\PassOptionsToPackage{unicode}{hyperref}
\PassOptionsToPackage{hyphens}{url}
\documentclass[superscriptaddress, reprint, showkeys]{revtex4-2}
\usepackage{amsmath,amssymb}
\usepackage{iftex}
\usepackage[dvipsnames]{xcolor}
\usepackage{subcaption}
\usepackage{booktabs,array}
\usepackage{multirow}
\usepackage{calc} 
\usepackage{etoolbox}
\usepackage{footnote}
\usepackage{graphicx}
\usepackage{bookmark}
\usepackage{babel}
\usepackage[normalem]{ulem}
\IfFileExists{xurl.sty}{\usepackage{xurl}}{} 
\usepackage{soul}
\usepackage{placeins}

\hypersetup{
  hidelinks,
  pdfcreator={LaTeX via pandoc}}
\usepackage{textgreek}
\usepackage{makecell}
\usepackage{multirow}
\usepackage{booktabs}
\usepackage{subcaption}
\usepackage{mathtools}
\usepackage{braket}
\usepackage{leftindex}
\date{\today}
\usepackage[nameinlink,capitalise]{cleveref}

\usepackage{stackengine}
\NewDocumentCommand{\leftstack}{ o m m m }{%
  \IfNoValueTF{#1}%
    {\leftindex^{\scriptstyle #3}_{\scriptstyle #4}{#2}}%
    {\leftindex[#1]^{\scriptstyle #3}_{\scriptstyle #4}{#2}}%
}

\begin{document}

\title{Machine-learning octet AB-type binary compounds across chemical space with domain knowledge of the interatomic bond}

\begin{abstract}
The prediction of the structural stability of octet $AB$-type binary compounds is a classical materials informatics problem. 
The challenge is to capture the relative stability of 4-fold coordinated atoms in zincblende ($\beta$-ZnS) structure and 6-fold coordinated atoms in rocksalt (NaCl) structure, modulated by charge transfer and atomic-size differences. 
Previous structure maps and machine-learning approaches used atomic features such as valence-electron count, ionization potential and atomic radii, using either physical intuition or symbolic regression. 
Here, we demonstrate that explicitly incorporating the domain knowledge of the interatomic bonds can significantly and systematically improve the prediction of $\beta$-ZnS/NaCl stability. 
We encode this bonding information through a coarse-grained representation of the local electronic structure obtained by a recursive solution of a tight-binding bond model. 
The underlying pairwise Hamiltonians are taken from downfolded eigenstates of density-functional theory calculations for diatomic molecules and thereby include  domain knowledge of the bond between specific $A-B$ pairs. 
The benefit of this description is demonstrated with an ensemble of independently trained Kernel Ridge or symbolic regression models combined with sequential feature selection. 
The obtained models are compared to a previous symbolic-regression model using the same set of \emph{ab initio} calculations for octet binaries as training data. 
We find a significant improvement in the prediction of the formation energy difference of $AB$ compounds as compared to previous works and demonstrate that an increasing amount of bond-informed recursion features improves the predictive accuracy.
\end{abstract}

\author{Rohan Kumar}
\email{rohan.kumar@rub.de}
\affiliation{Interdisciplinary Centre for Advanced Materials Simulation (ICAMS), Ruhr-Universit{\"a}t Bochum, 44801 Bochum, Germany}

\author{Mariano Forti}
\email{mariano.forti@rub.de}
\affiliation{Interdisciplinary Centre for Advanced Materials Simulation (ICAMS), Ruhr-Universit{\"a}t Bochum, 44801 Bochum, Germany}

\author{Aakash A. Naik}
\affiliation{Department of Materials Chemistry, 
Federal Institute for Materials Research and Testing, 12205 Berlin, Germany}
\affiliation{Institute of Condensed Matter Theory and Optics, Friedrich-Schiller-Universit{\"a}t Jena, 07743 Jena, Germany}

\author{Luca M. Ghiringhelli}
\affiliation{Scientific Computing Center, Karlsruhe Institute of Technology, Karlsruhe, Germany}

\author{Thomas Hammerschmidt}
\affiliation{Interdisciplinary Centre for Advanced Materials Simulation (ICAMS), Ruhr-Universit{\"a}t Bochum, 44801 Bochum, Germany}


\maketitle

\section{Introduction}

The quality of machine-learning models for predicting material properties depends critically on the features that are used to represent the material. At the atomistic level, the features need to capture geometric details of the crystal structures as well as chemical details of the constituent chemical elements.

A classical example for the inherent challenges of predicting the structural stability with ML approaches across chemical space is the case of octet binaries. These compounds have two elements and 8 valence electrons per formula unit. The chemical bonding in octet binaries ranges from highly ionic to covalent, crystallizing either in rocksalt or zincblende structures. Rocksalt crystal structures have 6-fold octahedral coordination with more ionic bonding while zincblende structures exhibit 4-fold tetrahedral coordination with covalent and directional bonding. 
Generally, a large electronegativity difference and a more ionic character of elements in octet binary compounds favours the rocksalt structure. The structural stability of the octet binaries is driven by atomic-size differences that determine lattice compatibility and by charge-transfer properties that determine the fulfillment of the octet rule. 

Pilania \emph{et al.}~\cite{pilania_classification_2015} classified ZB/RS structures in octet binary compounds based on dynamical charges.
They used a two-dimensional feature pair of the difference in Born effective charges between ZB and RS structures and Harrison’s bond polarity to train a support vector machine (SVM) classifier on 82 octet binary compounds.
This model reached $0.956$ average classification accuracy with excess Born effective charge ($\Delta Z_{\text{ex}}^{*}$) and pseudo-potential orbital radii ($r_{\sigma1}$), and a comparable $0.932$ with Born effective charge and bond polarity $(\Delta Z^{*},\alpha_p)$.
Their results demonstrate that small changes in dynamical charge and bond polarity reflect an atom’s response to its coordination and are powerful predictors of phase stability. 

Adutwum and Oliynyk ~\cite{Oliynyk-2016} extended this classification to a larger dataset of 706 octet binary compounds.
They used cluster resolution feature selection on an initial set of 56 elemental features and trained partial least-squares discriminant analysis (PLS-DA) and SVM with selected features to classify binary $AB$ compounds into seven prevalent crystal structures. Their SVM model also achieved a classification accuracy of 93.2\%. 

Independently of each other, Least Absolute Shrinkage and Selection Operator (LASSO) regression~\cite{Ghiringhelli-15} for \emph{ab initio} data and structure maps~\cite{Bialon-16} for experimental data showed that only three features derived from basic properties of the chemical elements, number of valence electrons, atomic radii, and electronegativity differences, are sufficient for a robust prediction of the ground state of octet binaries. 
In the work by Ghiringhelli {\it et al.}~\cite{Ghiringhelli-15}, energy differences between RS and ZB structures were computed for 82 octet binary materials. 
Using a LASSO-based feature selection on a diverse set of atomic descriptors, they identified a three dimensional descriptor that reached a cross validated root-mean-square error (RMSE) of 0.08 eV and a maximum absolute error (MaxAE) of 0.16 eV, averaged over 150 random leave-10\%-out splits.

An extension of these features based on general properties of the chemical elements is the representation of the electronic-structure of a particular compound in a particular crystal structure. 
Post-processing self-consistent DFT calculations leads to electronic fingerprints that range from binning the bandstructure and total density of states (DOS) to one-dimensional feature vectors~\cite{Isayev-15}, similarity coefficients between two total DOS~\cite{Kuban-2022} to atom-resolved DOS and bond-resolved crystal orbital indicators from projection onto an atomic orbital basis~\cite{Naik-2023}.
Attempts to avoid the computationally expensive self-consistent DFT calculations include one-hot encoding of the interatomic bonds as orbital-interaction matrix~\cite{Karamad-2020} and the inference of Fermi level, valence-band maximum and conduction-band minimum from non-self-consistent DFT calculations with pseudo-atomic orbitals~\cite{Zadoks-2024}. 

In this work, we demonstrate that chemistry-aware features of the atom-resolved DOS can also be obtained from coarse-grained electronic-structure calculations. 
We use a computationally inexpensive approach to compute the local DOS by recursion~\cite{Haydock-80-1,Haydock-80-2} with a tight-binding (TB) bond model. 
The local DOS is represented in terms of recursion coefficients~\cite{CryotLackmann-67, Ducastelle-70, Ducastelle-71} that capture the major part of the information for discriminating different local atomic environments already in the first few recursion levels~\cite{Ducastelle-70, Turchi-83, Bieber-83, Seiser-11-2, Hammerschmidt-16-2, jenke_electronic_2018}. 
These DOS features can directly be used to train ML models as shown in previous works~\cite{Sutton-19, Doesinger-25, forti2026data}. 
The performance of the resulting ML models can be further improved by introducing domain knowledge of the interatomic bond to the DOS features via TB Hamiltonians with bond-specific projections of DFT eigenstates to an atom-centered minimal basis~\cite{jenke_tight-binding_2021}.
Here, we apply this methodology to the case of binary $AB$-type octet binaries and compare the resulting chemistry-aware ML models to previous work using the same DFT dataset.

\section{Method}\label{sec:method}

\subsection{Domain knowledge of interatomic bond}\label{ssec:domainknowledge}

Atomic and composition based features are widely used in machine-learning property-prediction studies. 
However, atomic features lack fine and large structural details and long-range interactions which are critical for understanding chemical behaviour in many practical applications. 
Though DFT gives more accurate results, it has limits to scalability to large systems, due to the high computational cost. 

Here, we utilized recursion-based features to describe the interatomic interaction.
The recursion coefficients are computed with the \textit{BOPfox} software~\cite{hammerschmidt_bopfox_2019} on the basis of pairwise tight-binding Hamiltonians $H_{i\alpha j \beta}$ between atom $i$, orbital $\alpha$ and atom $j$, orbital $\beta$ in two-center approximation. 
Chemistry-specific Hamiltonian matrix elements $H_{i\alpha j \beta}$ are  taken from a
downfolding of DFT eigenstates of two-atomic molecules to an atom-centered minimal basis~\cite{jenke_tight-binding_2021}.
The resulting Hamiltonian is transformed to a tri-diagonal form using a Lanczos algorithm ~\cite{1950Lanczos} with corresponding recursion coefficients  $a_{n}$ and $b_{n}$~\cite{Aoki-93-2}: 

\begin{multline}\label{eqn:tri_dimensional_ham}
  \bra{u_{m}} H_{ij} \ket{u_{n}} = 
  \left(
  \begin{array}{ccccccc}
    a_{0}   & b_{1}   &         &          &         &        &        \\
    b_{1}   & a_{1}   & b_{2}   &          &         &        &        \\
            & b_{2}   & a_{2}   & b_{3}    &         &        &        \\
            &         & b_{3}   & a_{3}    & b_{4}   &        &        \\
            &         &         & \ddots   & \ddots  & \ddots &        \\
            &         &         &          & \ddots  & \ddots & \ddots
  \end{array}
  \right)
\end{multline}

\noindent where $\ket{u_i}$ are the Lanczos states generated from the orbital of interest.
Increasing order $n$ of the recursion coefficients corresponds to sampling an extended environment around atom $i$.

\begin{figure}[b]
    \centering
    \includegraphics[width=0.9\columnwidth]{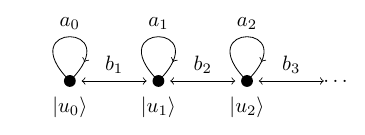}
    \caption{
    Graphical representation of the tri-diagonal Hamiltonian obtained by Lanczos recursion.
    }
    \label{fig:recursion_chain}
\end{figure}

The recursion coefficients are directly related to the local electronic DOS $n_{i\alpha}(E)$ of orbital $\alpha$ of atom $i$ by

\begin{equation*}
  n_{i\alpha}(E) = -\frac{1}{\pi} Im
  \left\{
    \vcenter{\hbox{$\cfrac{1}{
      E - a_{0}^{i\alpha} - \cfrac{(b^{i\alpha}_{1})^2}{
        E - a_{1}^{i\alpha} - \cfrac{(b^{i\alpha}_{2})^2}{\vdots}}}$}}
  \right\}
\end{equation*}

Here, the recursion coefficients for \textit{s} and \textit{p} orbitals are computed using one s and three p orbitals per atom.
Following the concept of reduced TB for $sp$-valent systems~\cite{Gehrmann-2015}, we use the average of the atomic recursion coefficients as atomic feature, 

  \begin{equation}
    a_n = ( a_n^{s} + 3 a_n ^{p} )/4
  \end{equation}

\noindent Descriptors resolving  $s$ and $p$ orbitals produce only a small improvement with respect to these averaged descriptors (see SI).

\subsection{Dataset description }\label{subsec:datasetdescription}

\begin{figure}[htbp!]
    \centering
    \includegraphics[width=\linewidth]{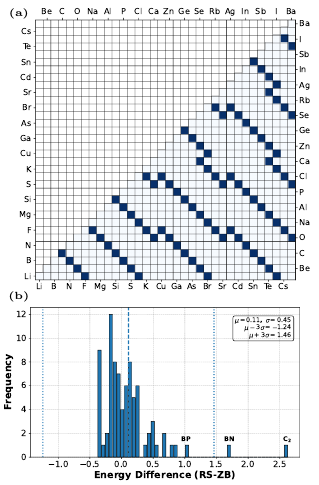}%
    \caption{
    (\subref{subfig:compoundcompos}) Matrix visualization of octet binary dataset (only one half of symmetric data is shown)
    (\subref{subfig:targetdistribution}) Distribution of the target variable.
   Broken and dotted vertical lines denote mean value and 3$\sigma$ range of the distribution, respectively.
    }%
    \label{fig:matrix_plot}%
    \phantomsubcaption\label{subfig:compoundcompos}%
    \phantomsubcaption\label{subfig:targetdistribution}%
\end{figure}

We used the same dataset of 82 octet binary compounds as in Ref.~\onlinecite{Ghiringhelli-15} to predict the formation energy difference $\Delta E_{AB}$ between rocksalt (RS) and zincblende/wurtzite (ZB/WZ) structures. 

\begin{equation}\label{eqn:target_property}
  \Delta E_{AB} = \Delta H_{F, AB}^{\textrm{RS}} - \Delta H_{F, AB}^{\textrm{ZB}}
\end{equation}

\noindent where $\Delta H_{F,AB}^{\varphi}$  is the formation enthalpy of the $AB$ compound in the $\varphi$-phase ($\varphi$ = RS or ZB).
A negative value of $\Delta E_{AB}$ indicates that a compound is more stable in RS structure. 
The dataset includes $AB$-type compounds from main-group elements, excluding transition metals to ensure consistent oxidation states, but certain post-transition elements are also included in a few compounds with fixed oxidation states.
The compounds included in the dataset are visualized in Figure~\ref{fig:matrix_plot}. 
Alkali metals, group 2 and group 13 elements like Li, Na, K, Mg, Al, and Ga are also present. 
This dataset covers a broad range of bonding character, from predominantly ionic halides and oxides to covalent nitrides, phosphides, and homoatomic group-IV compounds. 
The target variable defined in equation \eqref{eqn:target_property} ranges from $-0.38$~eV (SrTe) to $2.63$~eV (C$_2$). 

\subsection{Dataset featurization}\label{subsec:datasetfeaturization}

For a single compound $AB$, per-site scalar features can be calculated.
For building a robust feature space, we start our feature vector using fundamental atomic and electronic properties such as electronegativity, orbital radii ($s$, $p$, $d$), nuclear charge, valence electrons, HOMO and LUMO energy, ionization potential and many other properties of both elements.
Here, $A$ refers to the element with the lower first ionization potential, which provides a consistent and quantitative criterion for distinguishing the two elements across all binary compounds in the dataset. 
This feature set is referred to as $AtomF$ and its final shape is given by the concatenation of intrinsic material descriptors

\begin{equation}\label{eqn:AtomF_define}
  AtomF = \bigoplus _{q}  \lbrace q_A, q_B\rbrace
\end{equation}

\noindent where $\bigoplus$ indicates vector concatenation, in this case over all possible scalar features $q$.

Note that for a given chemical composition $AB$, both RS and ZB structures would yield the same $q_A$  and $q_B$, and then a unique $AtomF_{AB}$ as defined in equation \ref{eqn:AtomF_define} is sufficient as a feature of the compound although unable to differentiate atomic structures.

On top of these atomic properties, we computed recursion-based features.
Recursion coefficients are computed for $n \leq$ 8 (from the 1st to the 8th order), to capture complex bonding interactions.
Now, recursion coefficients depend not only on the chemical composition but also on atomic positions.
Then, we featurize a guess unit cell built from the structural prototypes scaled to an ideal mix of the atomic volumes according to Vegard's law \cite{vegard_konstitution_1921, denton_vegards_1991}. For a given crystal structure (i.e. RS or ZB), the recursion-based
features for a compound $AB$ can be calculated from comparison between the site-resolved recursion coefficients of the two inequivalent sites, 

\begin{equation}
    \begin{aligned}\label{eqn:structure_specific_features}
    Q^{max}_{\phi} &= \max \lbrace q_{\phi}^{ A},  q_{\phi}^{B} \rbrace \\
    Q^{ min }_{\phi} &= \min\lbrace q_{\phi}^{A}, q_{\phi}^{B} \rbrace \\
    Q^{ ave }_{\phi} &= \dfrac{q_{\phi}^{A} + q_{\phi}^{B}}{2}
    \end{aligned}
\end{equation}

\noindent where $\phi$ expands over the two structures, RS and ZB, and the scalar features are now given by the atomic recursion coefficients $a_n$ and $b_n$. 

In addition to this, the target property is the difference between the formation energies of the ZB and RS structures.
Hence, we can define chemistry specific features by defining  
per-compound features from comparison operators.
Specifically, we compute the difference ($\Delta$) and the ratio ($r$) between the per-structure feature from equation \ref{eqn:structure_specific_features}

\begin{equation}
\begin{aligned}\label{eqn:per_chemistry_specific_features}
Q^{f}_{\Delta}  &= Q^{f}_{RS} - Q^f _{ZB} \\
Q^f_r &= \dfrac{Q^{f}_{RS}}{Q^f_{ZB}} \\
\end{aligned}
\end{equation}

\noindent 
where $f$ expands over the $min$, $max$ and $ave$ operators defined above.

As the recursion coefficients can be calculated per site and later compared between RS and ZB structures as defined in equations \eqref{eqn:structure_specific_features} and  \eqref{eqn:per_chemistry_specific_features}, we define a feature vector $v_n$ as the concatenation of all those results, 

\begin{equation}
    {v_n} = \bigoplus_{f, O} 
    \lbrace
    (a_n)^{f}_{O},
    (b_n)^{f}_{O},
    \rbrace
\end{equation}

\noindent
where O expands over the comparison operators $\Delta$ and $r$.
In particular, a recursion feature vector of order $n$  would retain knowledge up to the $n$-th shell of neighbours. 
However, as RS and ZB are relatively simple structures, we expect that a ML model would benefit already from the first few recursion levels. 

Then, to capture the importance of domain knowledge on the structure retained in the features, we consider feature vectors with increasing order of the recursion coefficients, 

\begin{equation}
\begin{aligned}\label{eqn:definebopvectors}
  Rec_{N} =  v_1 \bigoplus \cdots \bigoplus v_N 
\end{aligned}
\end{equation}

These recursion-based descriptors capture how bonding characteristics (e.g., bond order, angular terms, and hybridization trends) shift between RS and ZB structures, giving the model insight into structural preferences at the electronic level.

\subsection{Data pre-processing} \label{subsec:datapreprocessing}

We used isolation forest~\cite{liu_isolation_2012} in two stages to identify outliers in our dataset. 
First, we used only the target variable together with the $AtomF$ descriptors. 

BN and C$_2$ appear as outliers due to exceptionally large positive $\Delta E$ as shown in \autoref{subfig:targetdistribution}, indicating a very strong preference for the ZB structure, which lies outside the typical range of the target variable. 
In the second stage, we expanded the feature set to include $Rec_N$ feature vector. 

\begin{figure}
    \centering
    \includegraphics[width=\linewidth]{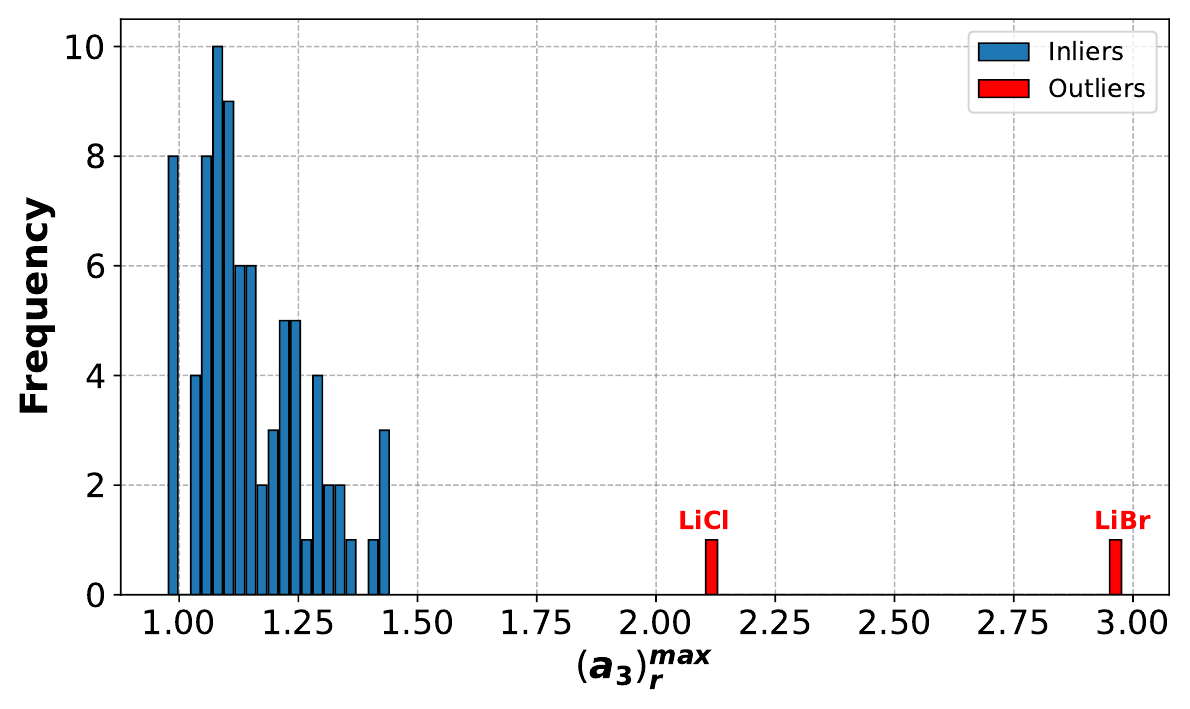}
  \caption{Distribution of $(a_3)^{max}_{r}$ showing the outliers.}
    \label{fig:outlierdetection}
\end{figure}

With this larger feature space, LiBr and LiCl behave as outliers, shown in \autoref{fig:outlierdetection}. 
Although their formation-energy differences are not extreme, they show unusually large shifts in certain recursion features (for example, in $(a_1)^{min}_{r}$,  $( b_2 )^{max}_{\Delta} $, $( a_3 )^{max}_{r} $).

By identifying these four outliers, we constructed two versions of the dataset: dataset 1 (D1) with all 82 data points and dataset 2 (D2) with 78 data points after removing all outliers.
This allows us to systematically evaluate the impact of these outliers on model learning and performance. 

For both datasets, multiple machine learning models are trained on different feature sets, starting with $AtomF$ feature vector and incorporating $Rec_{N}$ with increasingly larger $N$ and  $N_{max} = 8$.
This step-wise approach allows us to systematically evaluate and visualize how the inclusion of higher-order structural information influences ensemble model performance.

\subsection{Machine-learning architecture}\label{subsec:mlarchitecture}

\begin{figure*}[t]
    \centering
    \includegraphics[width=\textwidth]{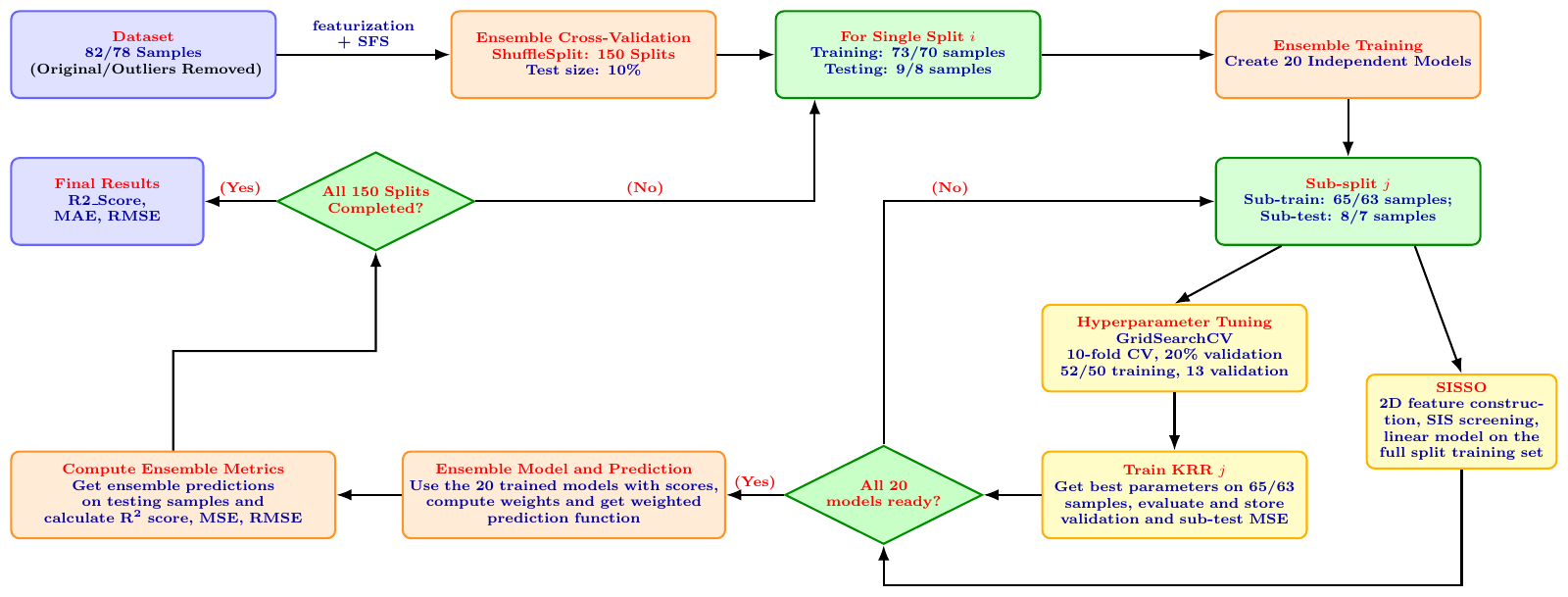}
    \caption{Flowchart showing model architecture and dataset utilization.}
    \label{fig:model_flowchart} 
\end{figure*}

We develop a supervised ML workflow as schematized in \autoref{fig:model_flowchart}, which shows how the small dataset is utilized within an ensemble model architecture.
Since our datasets contain either 82 or 78 samples only, we used an ensemble strategy to get more robust and generalized models. 
First, we perform sequential feature selection (SFS) as implemented in the \texttt{mlxtend} package~\cite{Raschka_mlxtend_2018} to reduce each feature set to a compact subset of descriptors.
This is a greedy forward search that adds, at each step, the feature that gives the largest improvement, with a conditional exclusion step that removes a previously added feature whenever this improves the score~\cite{pudil_floating_1994, ferri_comparative_1994}.
Candidate subsets are scored by the mean squared error of a KRR model with a polynomial kernel in a 10-fold cross-validation, and the subset size with the best score is retained.
The procedure is carried out separately for each feature set, so that the $AtomF$ model and each $AtomF+Rec_{N}$ model are built on their own selected features.
The SFS is performed once on the complete dataset, prior to the train-test splits described below, and the selected features are then kept fixed for all subsequent splits.

To obtain statistically robust estimates, we trained and evaluated 150 ensemble models using a repeated ShuffleSplit cross-validation scheme.
Specifically, the dataset was randomly split 150 times with a fixed random seed in each split, 90\% of the data was used for training and 10\% for testing, similar to the approach of \citeauthor{Ghiringhelli-15}~\cite{Ghiringhelli-15}.
Within each outer-split, the training set is shuffled and resampled 20 times into 90:10 sub-training and sub-test sets, and one base model is trained on each of these sub-splits.
A KRR and a SISSO model are independently trained for each sub-training set and evaluated on the sub-testing set, and their validation and sub-test errors are recorded.
The 20 KRR or SISSO models are combined into an ensemble model where the final estimation is obtained from a weighted average of the individual predictions, where the weights are derived from the recorded individual sub-test and validation performances (30\,\% and 70\,\% respectively) and converted into inverse MSE.
This ensemble approach helps to reduce the variance, improves the predictive reliability and achieves better performance than relying on a single model trained on the full dataset.
The ensemble model performance is evaluated on the 10\% held-out test set for each outer split.
Statistically robust estimates are obtained from the 150 resulting ensembles, with uncertainties determined from the standard deviation of the 150 ensemble models.

Some of the 150 outer splits produced ensembles with negative R\textsuperscript{2} and catastrophically large RMSE and MAE.
Those splits were ruled out when averaging the metrics reported below.

For each valid outer split, we calculated R\textsuperscript{2} score, mean absolute error (MAE), and root mean square error (RMSE) on both the training and held-out test sets. 
After computing these metrics across all splits, we evaluated the model performance in terms of mean and standard deviation of outer R\textsuperscript{2}, MAE, and RMSE for the training and testing samples. 
We also reported the maximum MAE and RMSE over all 150 folds, which is critical to understand the ensemble’s worst-case performance under different splits and reveal how poorly the model can perform under certain conditions, such as rare data patterns, outliers, or less representative feature sets. 

Finally, we also investigated the performance of three features from a previous study~\cite{Ghiringhelli-15} and used those metrics as a reference to compare recursion features within the ensemble model.

\subsection{SISSO}
The Sure Independence Screening and Sparsifying Operator (SISSO)~\cite{Ouyang-18} is a compressed-sensing method that identifies low-dimensional descriptors by applying mathematical operators to primary features and retaining the combination minimizing regression error.
For direct comparison to~\cite{Ghiringhelli-15}, we use the SISSO++ code~\cite{Purcel-23} to construct two-dimensional (2D) and three-dimensional (3D) descriptors from our $AtomF+Rec_{N}$ after the SFS selection.
These descriptors are linear combinations of 2 and 3 analytic terms respectively.
The mathematical operators used for obtaining SISSO models can be summarized as 

\begin{equation}
\begin{aligned}
\hat{H}^{(\mathrm{m})} \equiv {} &
\left\{
+,-,\times,\div,\exp,-\exp,\log,\vert\vert,
\right.\\
&
\left.
\vert\phi_1-\phi_2\vert,
\sqrt{},\sqrt[3]{},^{-1},^{2},^{3}
\right\}
\left[\phi_1,\phi_2\right].
\end{aligned}
\label{eq:sisso_operators}
\end{equation}

\noindent where $\mathrm{\phi_1}$ and $\mathrm{\phi_2}$ denote descriptors in the primary descriptor vector $\Phi_0$, i.e., SFS selected descriptors. The superscript $(m)$ enforces physical unit consistency, such that operations between descriptors with incompatible units are excluded. The hyperparameters used to obtain the SISSO descriptors are as follows. The selection of high-ranking constructed descriptors is controlled by the number of descriptors retained during the sure-independence screening step, ($n_{\text{sis}}$), which is set to 100. The number of residuals considered, ($n_{\text{residuals}}$), is set to 10. Finally, the complexity of the constructed descriptors is limited by restricting the maximum number of successive mathematical operations (the maximum rung) to 2.
Since a descriptor is constructed anew within every ensemble member, the ensemble selects a large number of distinct descriptors rather than a single one. The most frequently selected 3D descriptors, and the selection frequency of the primary features that enter them, are listed in the SI.

\section{Results and discussion}\label{sec:results}

First, we assessed the split-level stability of both KRR and SISSO ensembles by counting the outer splits whose held-out R\textsuperscript{2} is negative (i.e.\ worse than predicting the mean).
On the refined dataset D2 such invalid splits are rare, amounting to $0.1\%$ of the 150 splits for the KRR ensemble and $4.8\%$ for the 3D SISSO ensemble, whereas on the full dataset D1 they rise to $2.1\%$ and $13.7\%$, respectively.
The 2D SISSO ensemble fails on a comparable fraction of the splits ($5.1\%$ on D2 and $8.4\%$ on D1), so the higher descriptor dimension buys accuracy without buying stability.
The outlier LiBr is present in the majority of the failing splits on D1 ($86\%$ for the KRR ensemble, $71\%$ for the 2D SISSO ensemble).
A detailed per-compound analysis of these invalid splits is provided in the supplementary material.

\subsection{ML model for the full dataset}\label{subsec:mlmodels}

\begin{figure*}[ht]
    \centering
    \includegraphics[width=\textwidth]{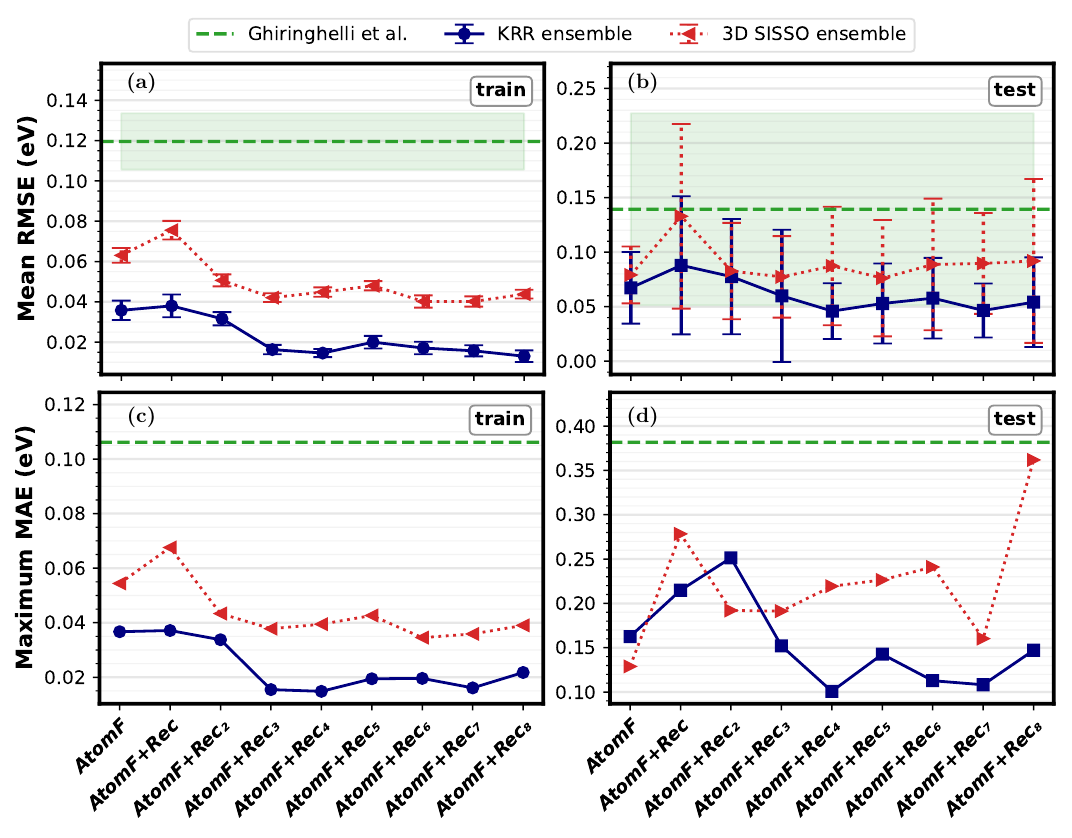}%
    \caption{
      Train and test performance of ensemble models evaluated on dataset D1 (82 samples).
    Mean RMSE~((\subref{subfig:conv82_meanrmse_train}) and (\subref{subfig:conv82_meanrmse_test})) and maximum MAE~((\subref{subfig:conv82_maxmae_train}) and (\subref{subfig:conv82_maxmae_test})) of the KRR ensemble are compared, on the same 150 splits, against the 3D SISSO ensemble and the reference model of Ghiringhelli \emph{et al.}~\cite{Ghiringhelli-15} (dashed green).
  Both ensemble curves are aggregated over the valid splits (R\textsuperscript{2}$\ge0$).
}%
    \label{fig:combined_convergence_82_samples}%
    \phantomsubcaption\label{subfig:conv82_meanrmse_train}%
    \phantomsubcaption\label{subfig:conv82_meanrmse_test}%
    \phantomsubcaption\label{subfig:conv82_maxmae_train}%
    \phantomsubcaption\label{subfig:conv82_maxmae_test}%
\end{figure*}

The results of our benchmark and model comparisons are shown in \autoref{fig:combined_convergence_82_samples}. 
On dataset D1, the ensemble model trained on the $AtomF$ set achieved better performance compared to the reference model.
Specifically, it reached a mean RMSE of $0.036\pm0.005$~eV on the training set and $0.067\pm0.033$~eV on the test set, whereas the reference model gives higher mean RMSE of $0.120\pm0.014$~eV for training and $0.139\pm0.088$~eV for testing.
The 3D SISSO ensemble, evaluated on the same 150 splits, also improves on the reference model, but remains less accurate than the KRR ensemble across the whole feature ladder: its mean test RMSE decreases from $0.079$~eV for the atomic descriptors to about $0.076$~eV at best.

As the recursion coefficients are incrementally added to $AtomF$ feature set, the model performance on the test set continued to improve.   
An ML model with the best performance is obtained when considering atomic features combined with recursion features up to the 4th order, which yields the lowest mean RMSE of $0.015\pm0.002$~eV for training and $0.046\pm0.026$~eV for testing.
Notably, this physically-informed feature set already outperforms the 3D SISSO ensemble.

Besides mean metrics, we also calculated maximum MAE and RMSE across 150 outer splits for all ensemble models which gives us insight into whether the model consistently performs well or if it still struggles with certain subsets of the data. 
The reference model is not able to perform well for all the splits and gives high maximum MAE of 0.106~eV for training and 0.382~eV for testing samples. 
However, as recursion features are added up to the 4th order, the model performance significantly improves as already observed in the mean RMSE metrics, reducing the maximum MAE to $0.015$~eV for training and $0.101$~eV for testing. This represents a major improvement in robustness compared to the 3D SISSO ensemble, whose maximum test MAE stays higher, in the range $0.13$--$0.36$~eV across the ladder. Further metrics are available in the supplementary material.

\subsection{ML model for the refined dataset}\label{subsec:mloptimizeddataset}

\begin{figure*}[ht]
    \centering
    \includegraphics[width=\textwidth]{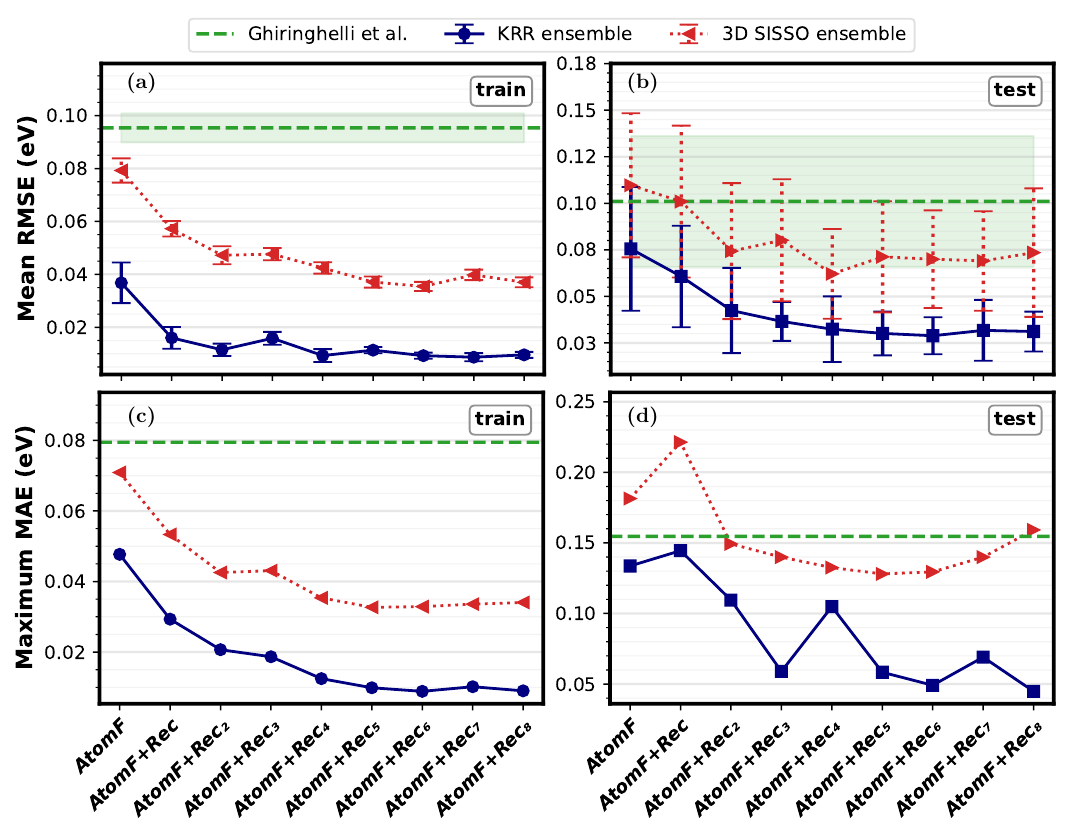}%
    \caption{Train and test performance of ensemble models with atomic and increasing order of recursive features, evaluated on dataset D2 (78 samples). Mean RMSE~((\subref{subfig:conv78_meanrmse_train}) and (\subref{subfig:conv78_meanrmse_test})) and maximum MAE~((\subref{subfig:conv78_maxmae_train}) and (\subref{subfig:conv78_maxmae_test})) of the KRR ensemble are compared, on the same 150 splits, against the 3D SISSO ensemble and the reference model of Ghiringhelli \emph{et al.}~\cite{Ghiringhelli-15} (dashed green). Both ensemble curves are aggregated over the valid splits (R\textsuperscript{2}$\ge0$). The 2D SISSO ensemble is left out here for clarity and is shown alongside the 3D one in the supplementary material.}%
    \label{fig:combined_convergence_78_samples}%
    \phantomsubcaption\label{subfig:conv78_meanrmse_train}%
    \phantomsubcaption\label{subfig:conv78_meanrmse_test}%
    \phantomsubcaption\label{subfig:conv78_maxmae_train}%
    \phantomsubcaption\label{subfig:conv78_maxmae_test}%
\end{figure*}

On D2 dataset, which excludes extreme target and descriptor values, the ensemble models produce lower prediction errors than on D1.
Here we show 3D SISSO ensembles in comparison to KRR ensembles, as they always perform better than the 2D SISSO ensembles, as detailed in the supplementary material.
We show in~\autoref{fig:combined_convergence_78_samples} that our feature set \textit{AtomF+Rec\textsubscript{N}}, even when using only atomic features, always outperforms the reference model and provides more accurate predictions. 
The best predictive performance was achieved using $AtomF$ along with recursion features up to the 6th order ($AtomF+Rec_{6}$), resulting in a mean RMSE of $0.009\pm0.001$~eV and $0.029\pm0.010$~eV for the train and test splits, respectively.
This ML model significantly exceeds the accuracy of the 3D SISSO ensemble, whose mean test RMSE decreases with recursion order but plateaus around $0.062$--$0.074$~eV, well above the KRR ensemble.
The $AtomF+Rec_{6}$ ensemble reaches this high-accuracy plateau without requiring additional symbolically discovered descriptors.

Notably, our ensemble model shows better accuracy in predictions even under the most adverse conditions among the 150 splits. 
The reference model shows poor stability across data subsets, with a high maximum test MAE of 0.155~eV. 
While the 3D SISSO ensemble is more stable than the reference model at higher recursion orders (maximum test MAE around $0.13$--$0.16$~eV), it remains significantly less robust than our model.
A consistent reduction in the training maximum MAE is observed, decreasing from 0.048~eV when using only atomic features to 0.009~eV upon incorporating $Rec_{6}$, beyond which the performance converges. 

The testing set shows a more non-monotonic variation in maximum MAE across recursion features, but overall it shows a clear improvement from 0.134~eV with only atomic features to 0.049~eV with recursion features up to the 6th order, and a slight further decrease to 0.045~eV when extended to 8th order.
However, the ensemble model incorporating recursion features up to the 8th order  ($AtomF+Rec_{8}$) yields a slightly higher mean RMSE, suggesting that the combination of atomic features with recursion coefficients up to the 6th order yields the most reliable and consistent performance across 150 splits. 
Other metrics are available in the supplementary material.
Furthermore, we conducted the same comparative analysis using 2D SISSO descriptors to check how much the symbolic baseline owes to the higher descriptor dimension.
The 3D descriptors are consistently the more accurate of the two, yet both remain well above the KRR ensemble on every feature set, so the conclusions drawn here do not depend on the dimension chosen for SISSO.
The corresponding metrics are provided in the supplementary material.

\subsection{Feature importance and selection frequency}

A detailed feature selection analysis on models trained both with D1 and D2 (with and  without outliers) highlights how atomic and recursion-based features contribute to prediction performance. 
Across both datasets, core atomic features such as ionization potential $(IP_A, IP_B)$, atomic orbital radii ($ r_A ^{p}, r_B ^{s} $ etc.), frontier orbital energies $(E_{\text{HOMO}}, E_{\text{LUMO}})$ were frequently selected from the simplest to the best-performing model, confirming their significant role in determining electronic structure and reactivity. 
Recursion features were also selected during feature selection process when they were included in the feature set.

In ensemble models trained on D2, both the diversity and number of selected recursion features increase progressively as recursion features from the 1st to 6th order are incrementally introduced. 
Lower order recursion features $( a_3 )^{ave}_{\Delta}$, $( a_1 )^{min}_{\Delta}$, $( b_2 )^{max}_{\Delta}$ were most frequently selected across several models because these features quantify nearest neighbour interactions such as bond lengths and bond order~\cite{hammerschmidt_bopfox_2019}, and effectively capture local coordination and bonding environment difference. 
As the recursion order increases, the features increasingly encode information from atoms in higher coordination shells, thereby capturing longer-range structural and chemical interactions. 

\begin{figure}[htb]
    \centering
    \includegraphics[width=0.5\textwidth]{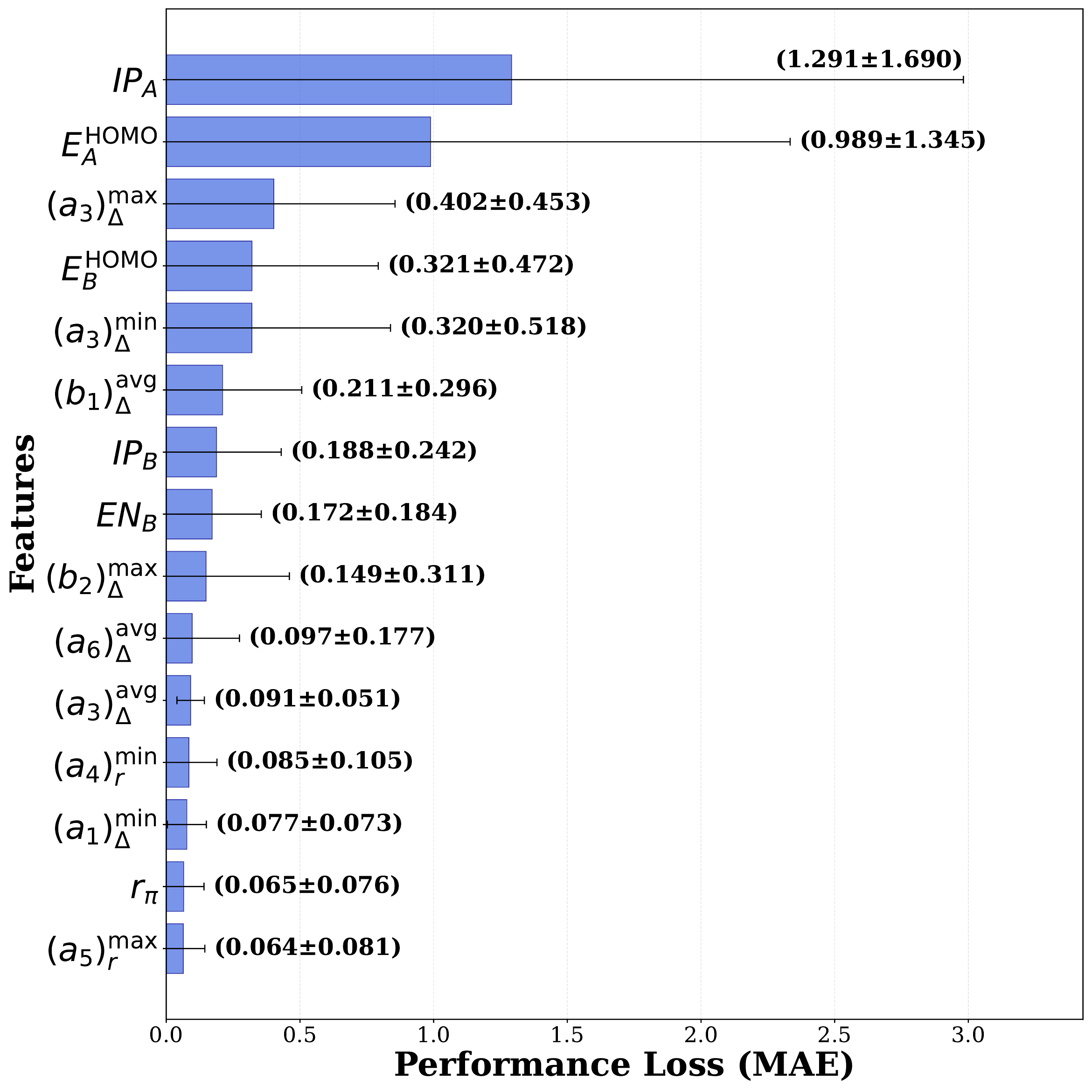}
    \caption{Permutation feature importance of best performing ensemble model. Bar length is the mean MAE increase when a feature is permuted and error bar showing the standard deviation over permutation repeats.}
    \label{fig:feature_importance}
  \end{figure}

As mentioned earlier, the ensemble model trained with atomic features and recursion features up to the 6th order achieved the best performance across all models. 
This model revealed a well balanced feature set that combines atomic, electronic and structural descriptors including frequently selected atomic features and at least one recursion feature from each order between the 1st and 6th orders. 
This feature set allows the model to capture both local and extended coordination environments effectively. 

SFS ignores the 7th order recursion coefficients and only retains one 8th order feature when considering $Rec_{7}$ and $Rec_{8}$ feature sets.
These results suggest that recursion orders 1–6 capture the majority of the relevant physics and achieve the optimal balance between accuracy and generalization, particularly when trained with clean, outlier-free data for simple octet binaries.

For models trained on D1 (82 samples, with outliers), more recursion features were selected overall, but their selection was less consistent across models, reflecting the higher descriptor variance introduced by the outliers. 
Removing the outliers (D2, 78 points) reduced this variance and stabilized the selected feature set. This indicates that atypical data can inflate the apparent importance of high order recursion features and blur what the model actually learns.
These convergence results from both datasets, with and without outliers, show that even the lowest-order recursion coefficients capture valuable structural information, most likely because the lower order recursion features directly correlate with bond strength differences between RS and ZB.

\begin{figure}[htb]
    \centering
    \includegraphics[width=0.5\textwidth]{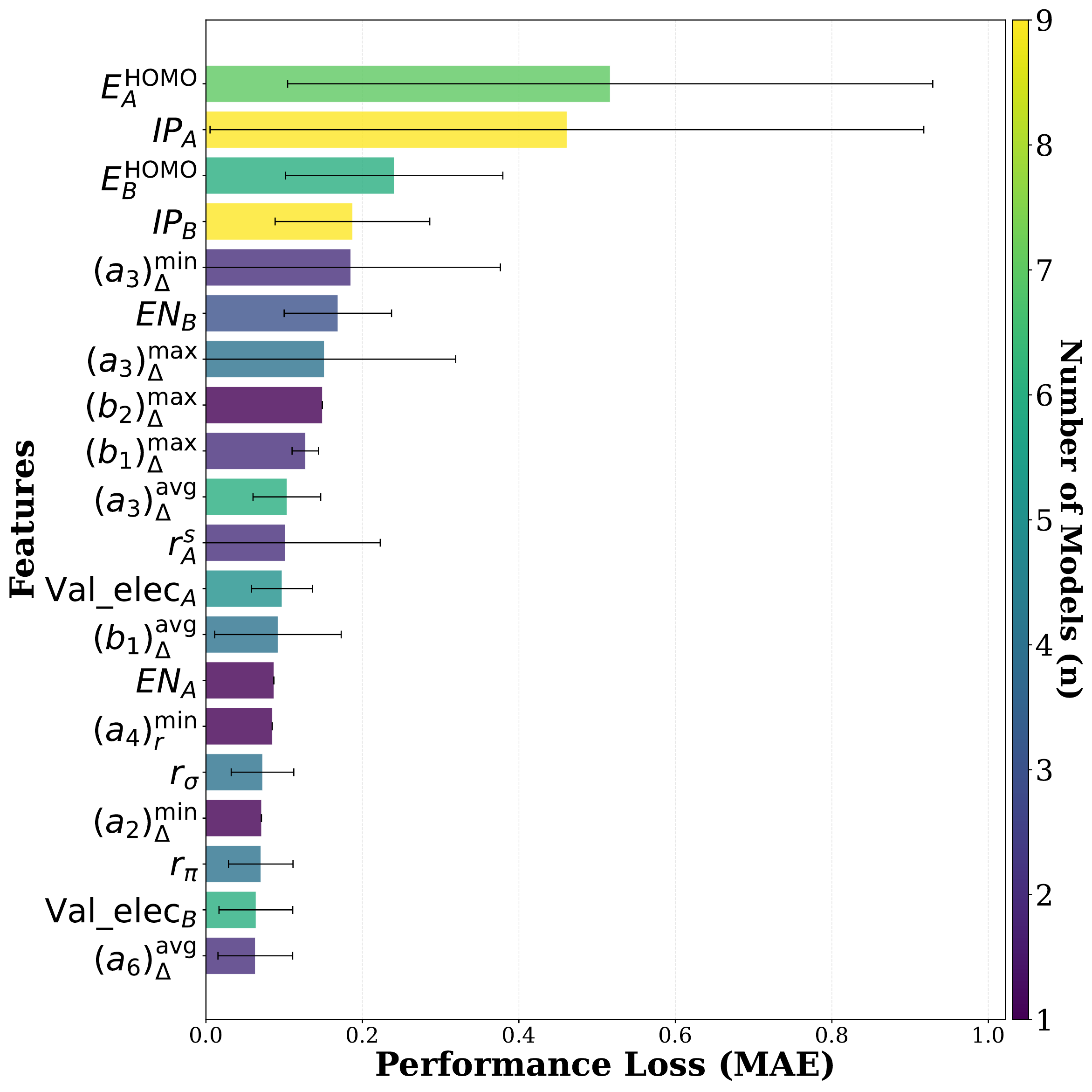}
    \caption{Permutation feature importance averaged across the ensemble models. Color bar encodes the selection frequency.}
    \label{fig:feature_importance_avg}
\end{figure}

\section{Conclusion}\label{sec:conclusion}

To understand the influence of features quantitatively, we evaluated permutation feature importance
(PFI)\cite{breiman2001random} using MAE as the metric, which measures the variation of MAE when a feature’s values are randomly modified. 
~\autoref{fig:feature_importance} shows the PFI of top 15 features of the best performing ensemble model. 
Atomic and electronic features such as ionization potential and HOMO energies have highest permutation importance, confirming that the model captures fundamental electronic drivers of stable structure. 
Several recursion features, particularly from the 3rd and 6th orders, also show significant increase in MAE during PFI analysis, confirming their contribution to the model accuracy.
Notably, the model selects average, maximum and minimum values of site resolved recursion descriptors, indicating that both overall bonding environment and differences between local environments pf A and B sites contribute to the prediction of structural stability.

The problem of the relative stability of the RS and ZB octet binaries across chemical space has been revisited from an improved ML point of view. 
We propose an ML workflow based on a physically informed feature set that combines fundamental atomic properties with recursion-based descriptors derived from coarse-grained electronic-structure calculations in comparison with SISSO symbolic regression. 
The ensemble model incorporating recursion-based features alongside atomic features (\textit{AtomF}) yields substantially lower errors than the reference model trained on the three features of Ghiringhelli \emph{et al.}~\cite{Ghiringhelli-15}, confirming the added predictive power of the physics-informed recursion-based descriptors. 
We further benchmarked against descriptors identified by 2D and 3D SISSO, where the algorithm was provided with the same atomic features alongside differences and ratios of recursion features to construct compact nonlinear symbolic descriptors. 
While SISSO-derived features alone improve over the reference model, the gains remain comparable to those achieved by SFS applied directly to the  recursion-based feature set. 
This confirms that the predictive improvement is primarily driven by the  recursion-based descriptors, with SISSO providing a complementary but secondary route through nonlinear symbolic combinations of the same underlying features.

We also demonstrate that features based on coarse-grained electronic-structure calculations with TB Hamiltonians yield atom-resolved features that lead to improved performance of ML models for structural-stability predictions. 
The representation of the atomic bonds in terms of recursion coefficients leads to rapid convergence of the ML accuracy with increasing recursion levels corresponding to  increasing farsightedness of the features and increasing level of detail of the DOS shape. 
The recursion coefficients are furthermore equipped with domain-knowledge of the interatomic bond via utilising chemistry-specific pairwise TB Hamiltonians. 

Our ML prediction of the structural stability of octet binary compounds across chemical space shows a significant improvement over previous works. 
The maximum absolute error and the mean RMSE across 150 random data splits were significantly decreased from 0.160~eV to 0.049~eV and from 0.080~eV to 0.029~eV, respectively. 
The resulting chemistry-aware features lead to ML models with remarkably high robustness given the broad variation of chemical compositions in a very small dataset. 
The significance of the chemistry-aware DOS features in comparison to general properties of chemical elements is verified by their high ranks in a feature importance analysis.

\section*{Data availability} 
The datasets and Jupyter notebooks used in this study are hosted in a dedicated repository and will be made available to reviewers for the evaluation process. 
They will be made fully public upon acceptance of the manuscript. 

\section*{Acknowledgements} 
RK, MF and TH acknowledge funding of this work by the German Deutsche Forschungsgemeinschaft (DFG No. 505643559 and 535248809). 
LMG and TH acknowledge funding from the German Research Foundation (DFG) through the NFDI consortium ``FAIRmat'', project 460197019.

\bibliography{main.bib}

\end{document}


\title{Supplementary material for article ``Machine-learning octet $AB$-type binary compounds across chemical space with domain knowledge of the interatomic bond''}

\author{Rohan Kumar}
\email{rohan.kumar@icams.rub.de}
\affiliation{Interdisciplinary Centre for Advanced Materials Simulation (ICAMS), Ruhr-Universit{\"a}t Bochum, 44801 Bochum, Germany}

\author{Mariano Forti}
\email{mariano.forti@icams.rub.de}
\affiliation{Interdisciplinary Centre for Advanced Materials Simulation (ICAMS), Ruhr-Universit{\"a}t Bochum, 44801 Bochum, Germany}

\author{Aakash A. Naik}
\affiliation{Department of Materials Chemistry, 
Federal Institute for Materials Research and Testing, 12205 Berlin, Germany}
\affiliation{Institute of Condensed Matter Theory and Optics, Friedrich-Schiller-Universit{\"a}t Jena, 07743 Jena, Germany}

\author{Luca M. Ghiringhelli}
\affiliation{Scientific Computing Center, Karlsruhe Institute of Technology, Karlsruhe, Germany}


\author{Thomas Hammerschmidt}
\email{thomas.hammerschmidt@icams.rub.de}
\affiliation{Interdisciplinary Centre for Advanced Materials Simulation (ICAMS), Ruhr-Universit{\"a}t Bochum, 44801 Bochum, Germany}

\maketitle

\renewcommand{\thefigure}{S\arabic{figure}}
\renewcommand{\thetable}{S\arabic{table}}

\section{Description of atomic features}
\begin{table}[!hbp]
    \centering
    \caption{Atomic descriptors used in this work}
    \includegraphics[width=\textwidth]{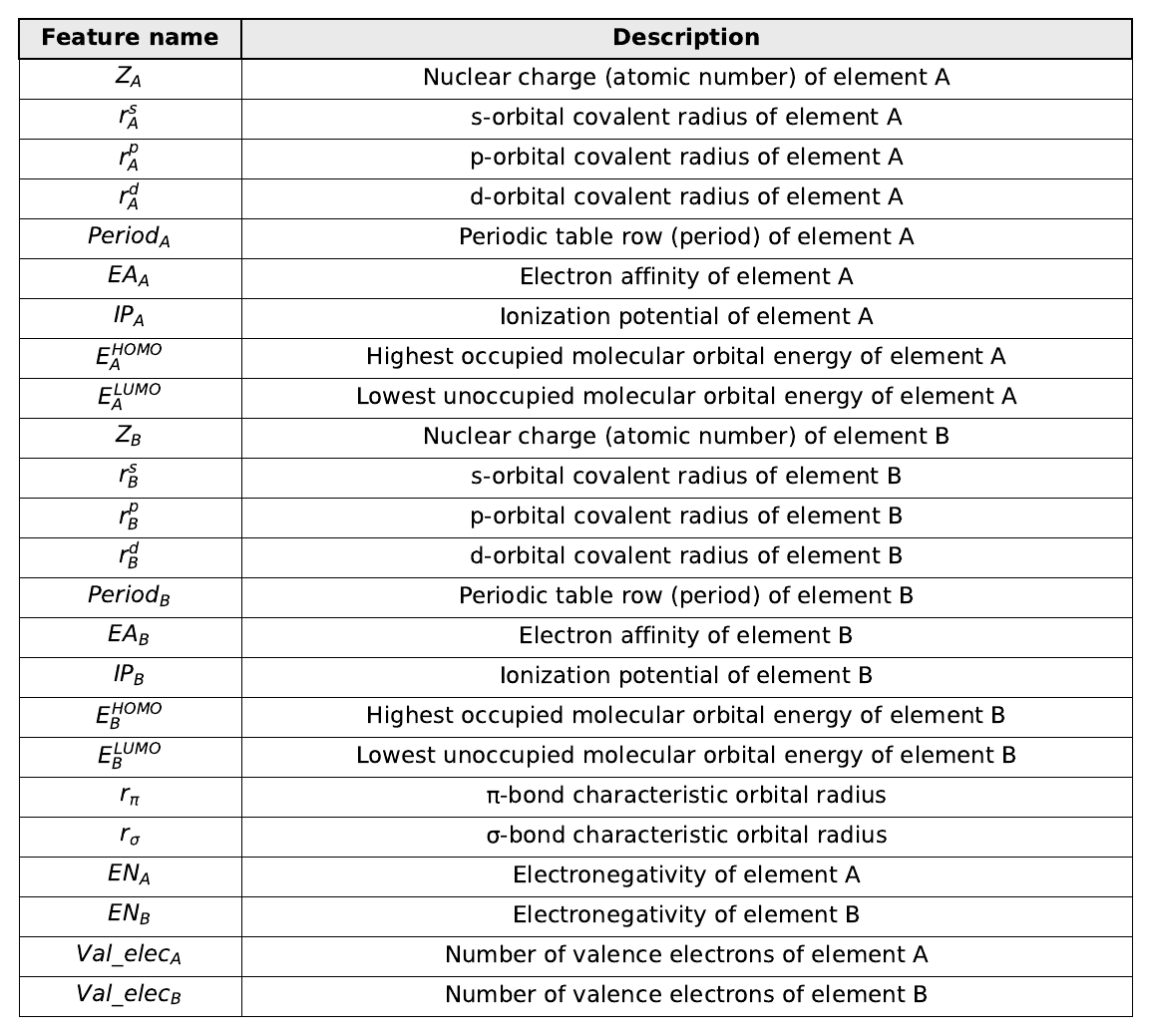}
\end{table}

\section{Performance metrics of the ensemble models and the 2D and 3D SISSO ensembles}

\begin{figure}[!hbp]
    \centering
    \includegraphics[width=\textwidth]{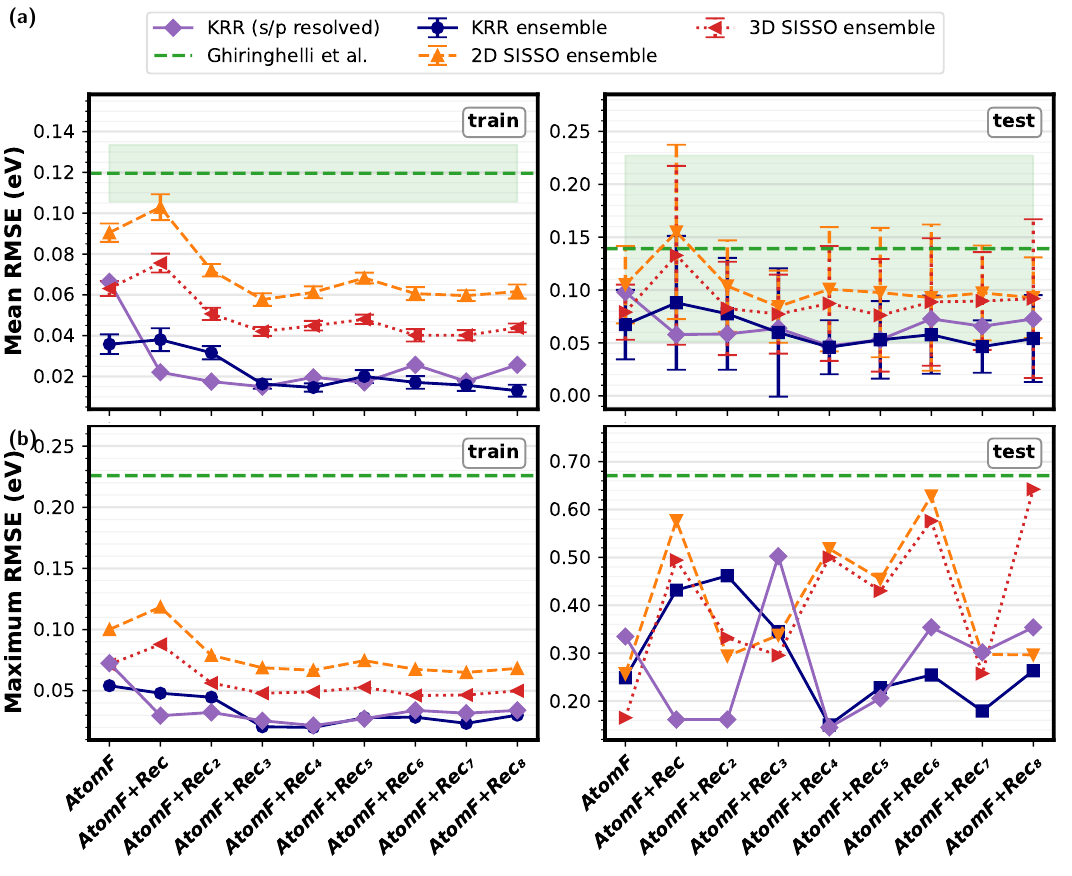}%
    \caption{Mean~(\subref{subfig:rmse82_mean}) and maximum~(\subref{subfig:rmse82_max}) RMSE across the 150 splits for dataset D1 (82 samples), with the train panel on the left and the test panel on the right. The KRR ensemble is compared against the 2D and 3D SISSO ensembles and the orbital-resolved ($s/p$) KRR ensemble (purple, no error bars).}%
    \label{fig:metric_rmse_82}%
    \phantomsubcaption\label{subfig:rmse82_mean}%
    \phantomsubcaption\label{subfig:rmse82_max}
\end{figure}

\begin{figure}[!hbp]
    \centering
    \includegraphics[width=\textwidth]{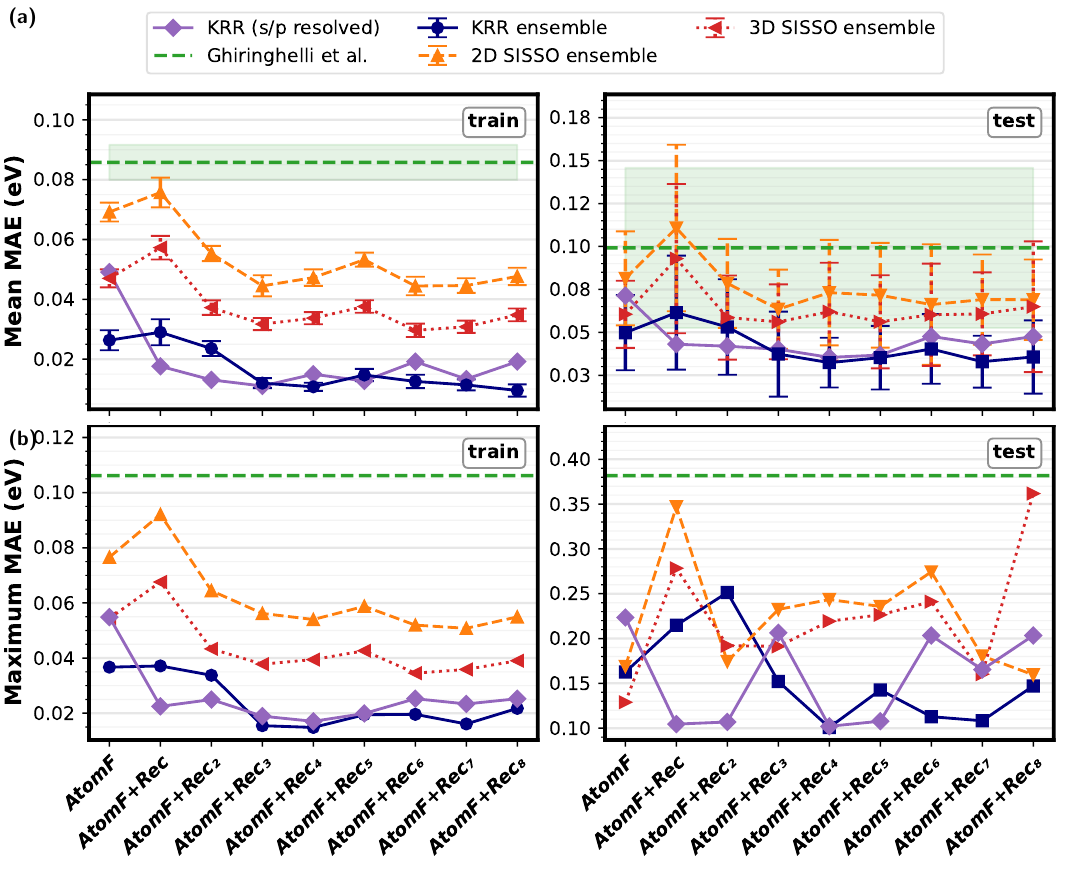}%
    \caption{Mean~(\subref{subfig:mae82_mean}) and maximum~(\subref{subfig:mae82_max}) MAE across the 150 splits for dataset D1 (82 samples), train (left) and test (right). Curves as in Figure~\ref{fig:metric_rmse_82}.}%
    \label{fig:metric_mae_82}%
    \phantomsubcaption\label{subfig:mae82_mean}%
    \phantomsubcaption\label{subfig:mae82_max}
\end{figure}

\begin{figure}[!hbp]
    \centering
    \includegraphics[width=\textwidth]{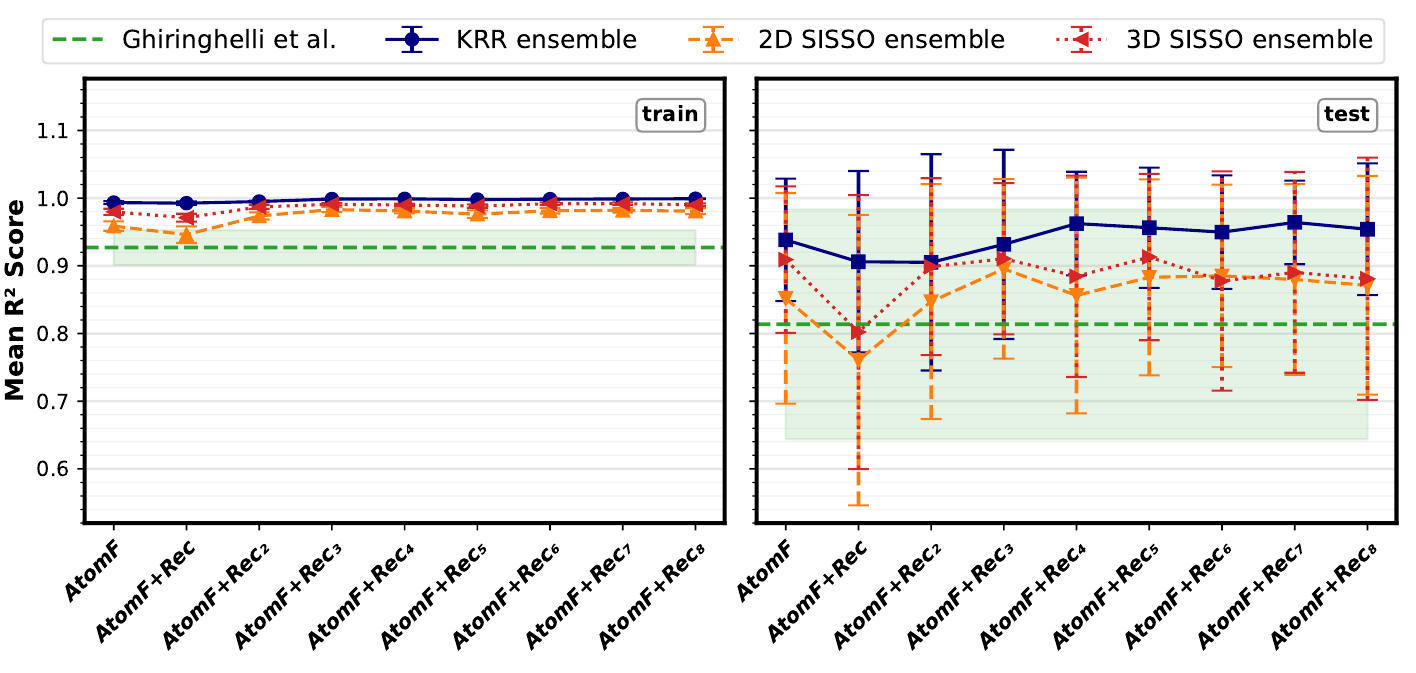}
    \caption{Mean $R^2$ score across the 150 splits for dataset D1 (82 samples), train (left) and test (right). Curves as in Figure~\ref{fig:metric_rmse_82}.}
    \label{fig:metric_r2_82}
\end{figure}

\begin{figure}[!hbp]
    \centering
    \includegraphics[width=\textwidth]{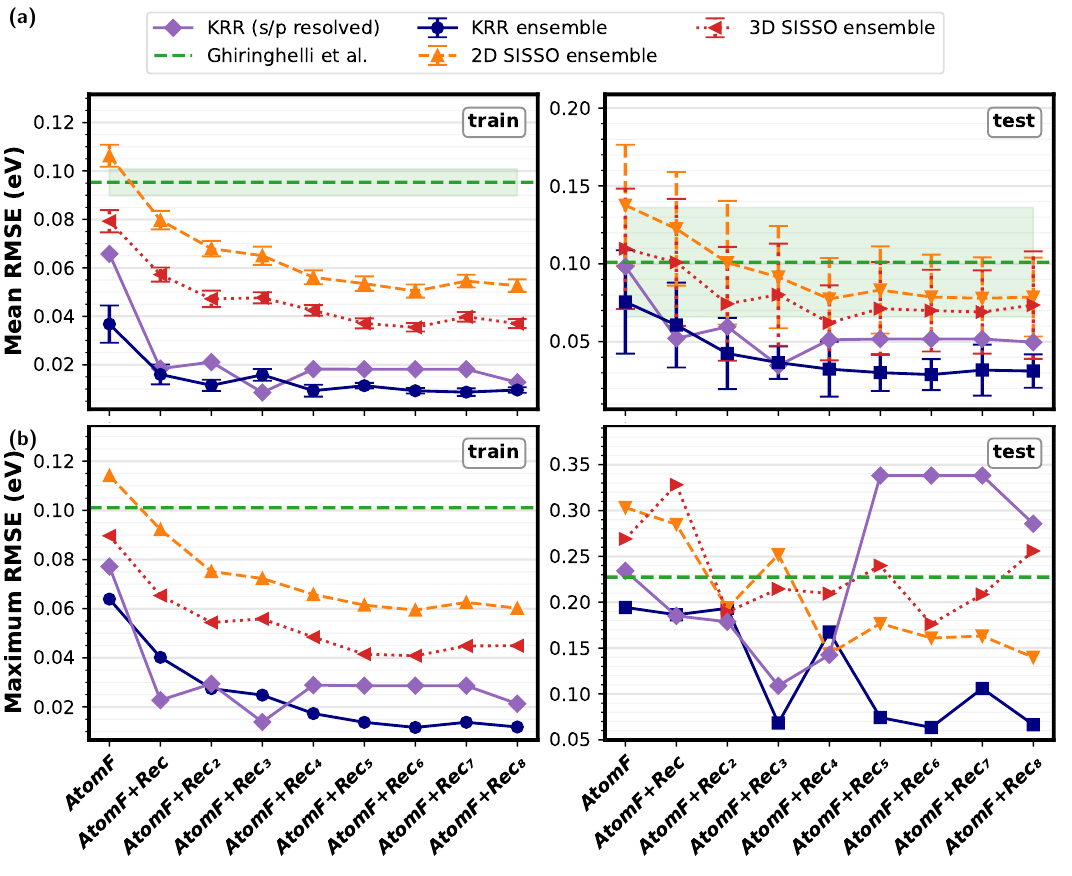}%
    \caption{Mean~(\subref{subfig:rmse78_mean}) and maximum~(\subref{subfig:rmse78_max}) RMSE across the 150 splits for dataset D2 (78 samples), train (left) and test (right). Curves as in Figure~\ref{fig:metric_rmse_82}.}%
    \label{fig:metric_rmse_78}%
    \phantomsubcaption\label{subfig:rmse78_mean}%
    \phantomsubcaption\label{subfig:rmse78_max}
\end{figure}

\begin{figure}[!hbp]
    \centering
    \includegraphics[width=\textwidth]{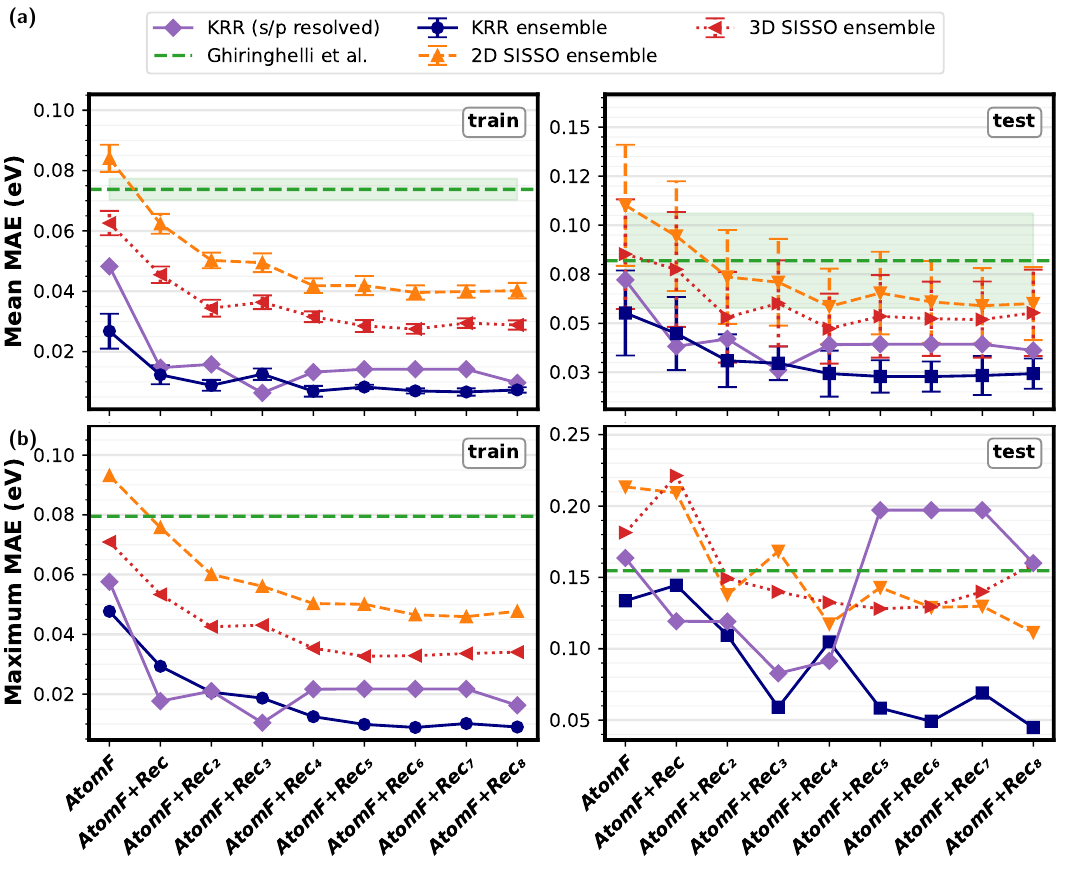}%
    \caption{Mean~(\subref{subfig:mae78_mean}) and maximum~(\subref{subfig:mae78_max}) MAE across the 150 splits for dataset D2 (78 samples), train (left) and test (right). Curves as in Figure~\ref{fig:metric_rmse_82}.}%
    \label{fig:metric_mae_78}%
    \phantomsubcaption\label{subfig:mae78_mean}%
    \phantomsubcaption\label{subfig:mae78_max}
\end{figure}

\begin{figure}[!hbp]
    \centering
    \includegraphics[width=\textwidth]{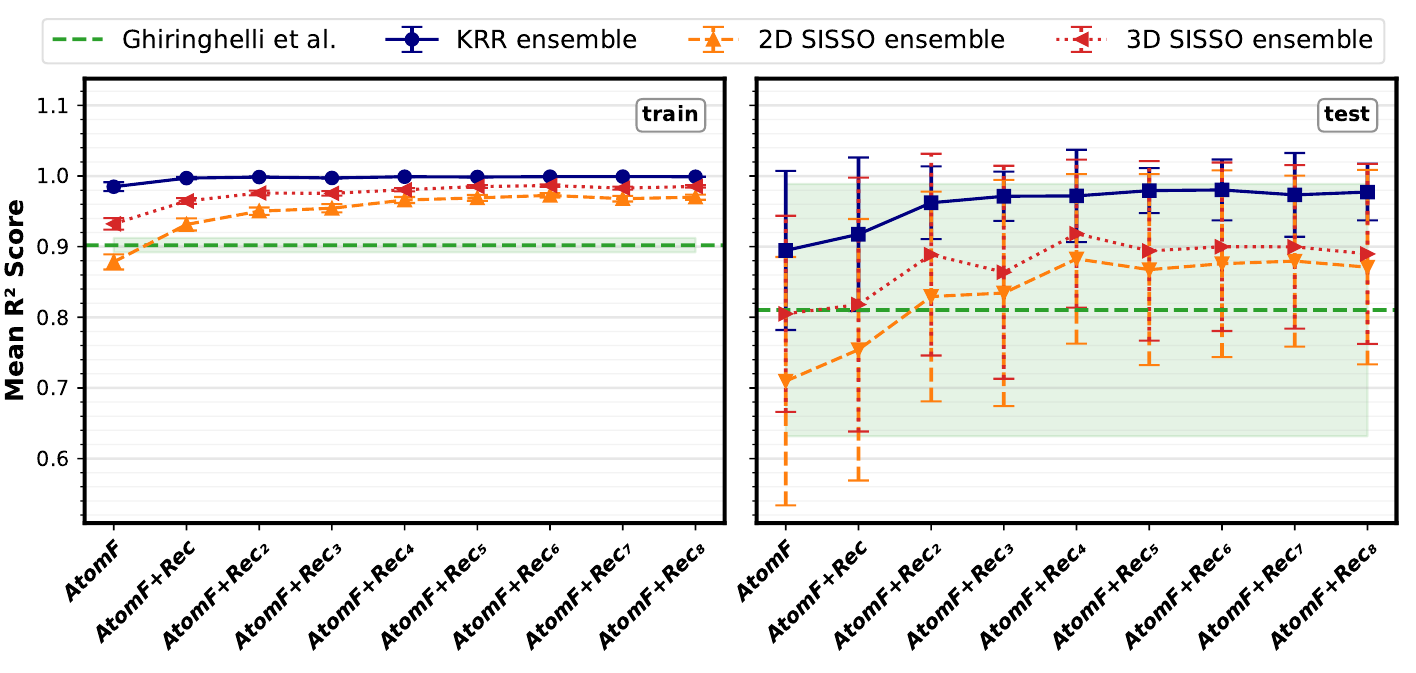}
    \caption{Mean $R^2$ score across the 150 splits for dataset D2 (78 samples), train (left) and test (right). Curves as in Figure~\ref{fig:metric_rmse_82}.}
    \label{fig:metric_r2_78}
\end{figure}

This section reports the full set of performance and stability metrics across the 150 outer splits, comparing the KRR ensemble against both the 2D and the 3D SISSO ensembles, together with an orbital-resolved ($s/p$) KRR ensemble.
The main manuscript shows only the mean RMSE and the maximum MAE, and against the 3D SISSO ensemble alone, since it is the more accurate of the two symbolic baselines and a single SISSO curve keeps those figures readable.
Here all the metrics are presented.
Figures \ref{fig:metric_rmse_82} to \ref{fig:metric_r2_78} stack the mean and the maximum over the 150 splits (bottom row), with the train panel on the left and the test panel on the right.
The $R^2$ score is shown as its mean only, since its maximum is not informative.
On every panel the orbital-resolved ($s/p$) KRR ensemble is drawn as a purple line without error bars.

The 3D descriptors give the lower mean test RMSE on every feature set of both datasets, by a margin that reaches about $0.03$~eV for the atomic descriptors and shrinks to about $0.01$~eV at the best-performing feature sets, but they are not the more stable ones: on D1 the fraction of invalid splits grows from $8.4\%$ in 2D to $13.7\%$ in 3D, whereas on D2 the two are comparable ($5.1\%$ against $4.8\%$).
Both remain well above the KRR ensemble throughout, so the comparison drawn in the main text does not depend on the SISSO dimension.

\clearpage
\section{Invalid splits and outlier co-occurrence}

For a fraction of the 150 outer splits, the held-out test subset yields a negative coefficient of determination
($R^2<0$), i.e.\ the ensemble performs worse than a constant predictor; we call these \emph{invalid} (failing) splits.
Because the KRR ensemble and the 2D SISSO ensemble share the identical split protocol, every split contains the same test compounds in both methods, enabling a direct compound-level attribution of the failures.

\autoref{tab:failed_folds_sisso} lists the number of failing test splits for each feature set. 
Pooled over the nine feature sets \textit{AtomF} to \textit{AtomF+Rec\textsubscript{8}}, the failing fraction is small on the refined dataset D2, where the KRR ensemble is essentially always valid ($0.1\%$ failing) while the 2D SISSO ensemble fails on $5.1\%$ of the splits. 
On the full dataset D1, which retains the four outliers (BN, C$_2$, LiBr, LiCl) identified by the isolation forest, the failing fraction rises to $2.1\%$ (KRR) and $8.4\%$ (SISSO).

\begin{table}[!htbp]
  \centering
  \caption{\protect\label{tab:failingD1splits} 
    Number of failed test splits (test $R^{2}<0$) out of 150, for the 2D SISSO
  ensemble and the KRR ensemble, on datasets D1 (82 compounds, outliers included) and
  D2 (78 compounds, outliers removed). SISSO fails on roughly four times as many splits
  as KRR on D1, and removing the outliers does not cure it.}
  \label{tab:failed_folds_sisso}
  \begin{ruledtabular}
  \begin{tabular}{llcccccccccc}
    Method & Dataset & $\mathit{AtomF}$ & $+\mathit{Rec}_{1}$ & $+\mathit{Rec}_{2}$ & $+\mathit{Rec}_{3}$ & $+\mathit{Rec}_{4}$ & $+\mathit{Rec}_{5}$ & $+\mathit{Rec}_{6}$ & $+\mathit{Rec}_{7}$ & $+\mathit{Rec}_{8}$ & Total \\
    \colrule
    \multirow[t]{2}{*}{SISSO 2D}   & D1 & 0 & \textbf{34} & 2 & 5 & 16 & 7 & 20 & 10 & 20 & \textbf{114} \\
                                  & D2 & 8 & 10 & 4 & \textbf{20} & 2 & 2 & 3 & 2 & 18 & \textbf{69} \\
    \colrule
    \multirow[t]{2}{*}{KRR} & D1 & 2 & 14 & 3 & 2 & 0 & 4 & 2 & 0 & 1 & 28 \\
                                  & D2 & 1 & 0 & 0 & 0 & 0 & 0 & 0 & 0 & 0 & \textbf{1} \\
  \end{tabular}
  \end{ruledtabular}
\end{table}

\begin{figure}[htb]
    \centering
    \includegraphics[width=0.7\textwidth]{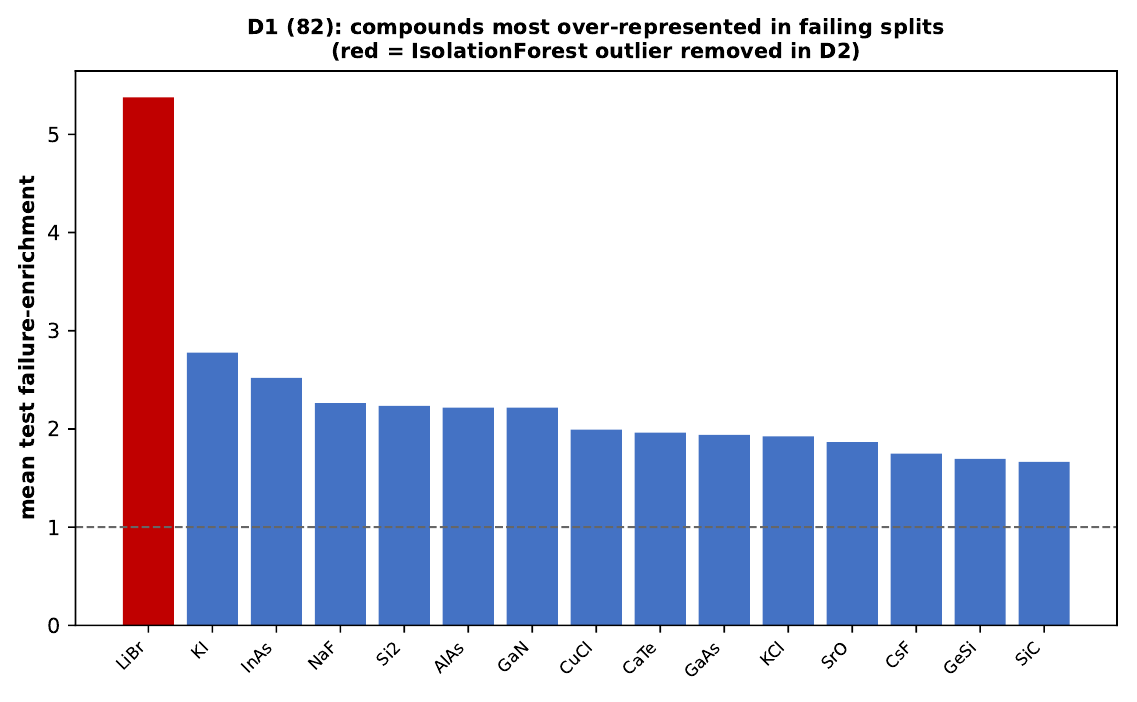}
    \caption{Compounds most over-represented in the failing test splits of dataset D1 (mean failure enrichment across feature sets). Red bars mark the isolation-forest outliers; the dashed line is the no-enrichment baseline ($E=1$).}
    \label{fig:invalid_enrichment}
\end{figure}

To identify which compounds drive the failures, we define for each compound $c$ the failure enrichment 

\begin{equation}
E(c)=P(c\in\mathrm{test}\mid\mathrm{split\ failed})/P(c\in\mathrm{test})
\end{equation}

\noindent values $E(c) \gg 1 $ indicate over-representation among the failing splits (\autoref{fig:invalid_enrichment}). 
On D1 the outlier LiBr dominates, appearing in $86\%$ (KRR) and $71\%$ (2D SISSO) of all failing splits.

\begin{figure}[htb]
    \centering
    \includegraphics[width=\textwidth]{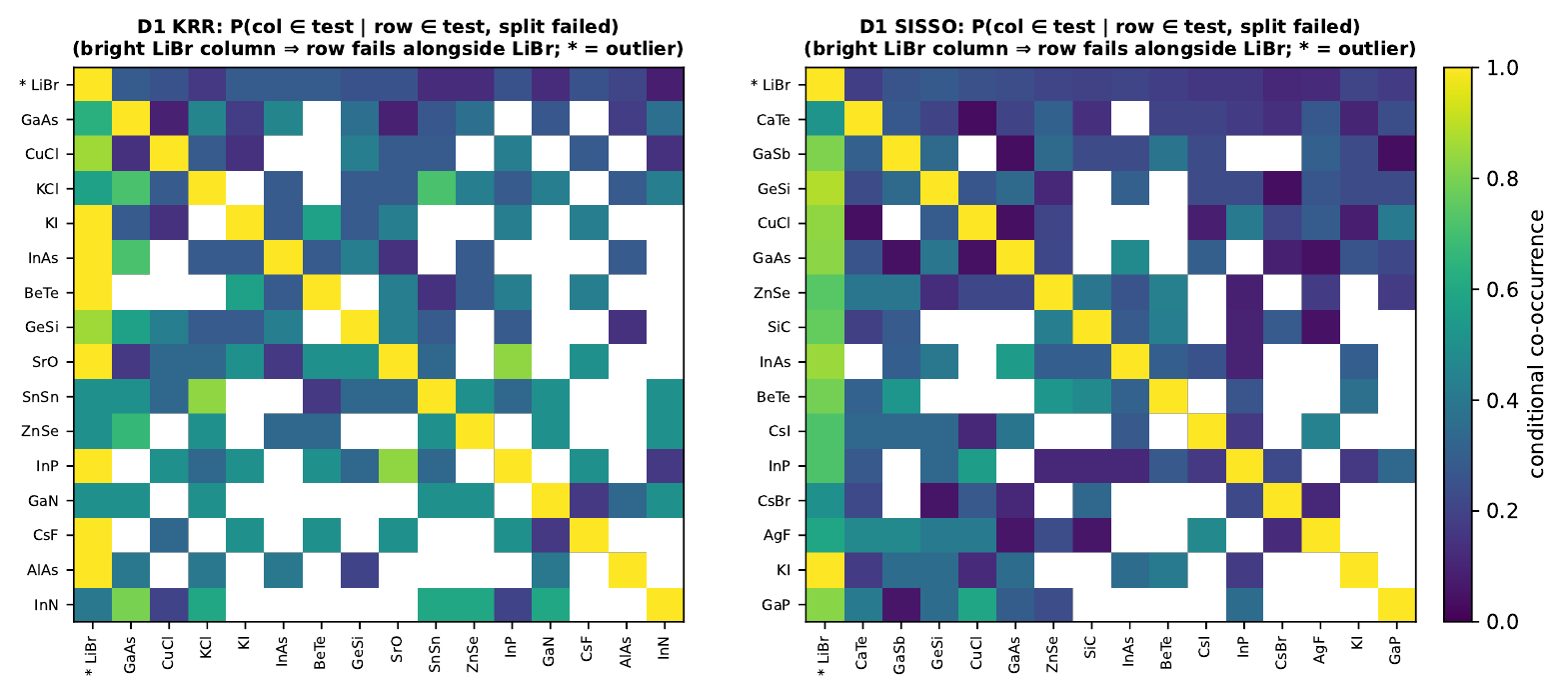}
    \caption{Conditional co-occurrence of compounds in the failing test splits of dataset D1, $P(\text{column}\in\mathrm{test}\mid\text{row}\in\mathrm{test},\ \mathrm{split\ failed})$, for the KRR ensemble (left) and 2D SISSO ensemble (right). A bright leftmost (LiBr) column identifies rows that fail predominantly in the presence of the outlier LiBr (guilt by association); asterisks mark the outliers.}
    \label{fig:invalid_cooccurrence}
\end{figure}

Marginal enrichment alone cannot separate a compound that fails \emph{independently} from one that merely shares its test split with LiBr. 
We therefore recompute the enrichment restricted to the splits in which LiBr is \emph{absent} from the test set. 
\autoref{fig:invalid_cooccurrence} shows the conditional co-occurrence $P(\text{column}\in\mathrm{test}\mid\text{row}\in\mathrm{test},\ \mathrm{split\ failed})$: a bright LiBr column marks compounds that fail only in the company of LiBr. 
Under the LiBr-absent control, compounds such as GaN, GaAs and KCl retain a high enrichment and are therefore genuine, independent sources of instability, whereas KI, InAs, AlAs, CsF, SrO and NaF lose essentially all enrichment: they fail only alongside LiBr and cannot be blamed independently of it.

\clearpage
\section{SISSO descriptors}

Every SISSO++ run returns a rank-2 (2D) descriptor, that is, a linear model built from two analytic terms. Because a descriptor is constructed anew inside each ensemble member, the nine feature sets, the 150 outer splits and the 20 members per split give $9\times150\times20\times2=54\,000$ term selections for each dataset. The ensemble therefore does not produce a single descriptor: pooled over all feature sets, $3\,555$ (D1) and $3\,778$ (D2) distinct terms are selected, and even the most frequent one accounts for only about $4\%$ of all selections. This spread is the descriptor-level counterpart of the split-to-split instability of the SISSO ensemble reported above.

\begin{figure}[htb]
    \centering
    \includegraphics[width=\textwidth]{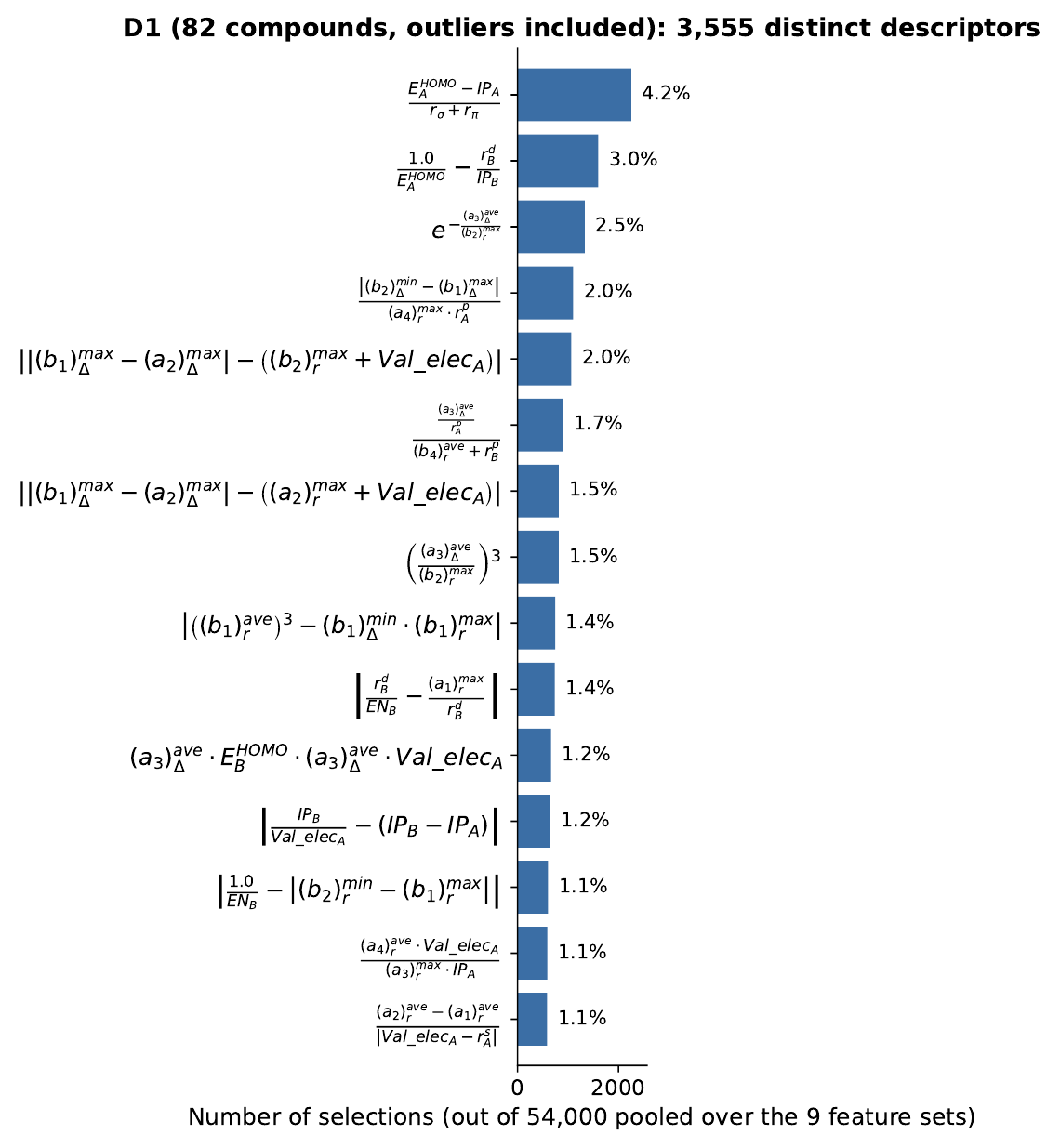}
    \caption{Globally most frequently selected 2D SISSO descriptor terms for dataset D1 (82 compounds, outliers included). The selections are pooled over all nine feature sets ($9\times150\times20\times2 = 54\,000$ term selections); the fifteen most frequent terms are shown, with the bar length the number of selections and the annotation its global share. The title gives the total number of distinct terms selected. Even the most frequent descriptor is chosen in only about $4\%$ of all selections, so the ensemble does not converge on a single descriptor.}
    \label{fig:sisso_desc_freq_D1}
\end{figure}

\begin{figure}[htb]
    \centering
    \includegraphics[width=\textwidth]{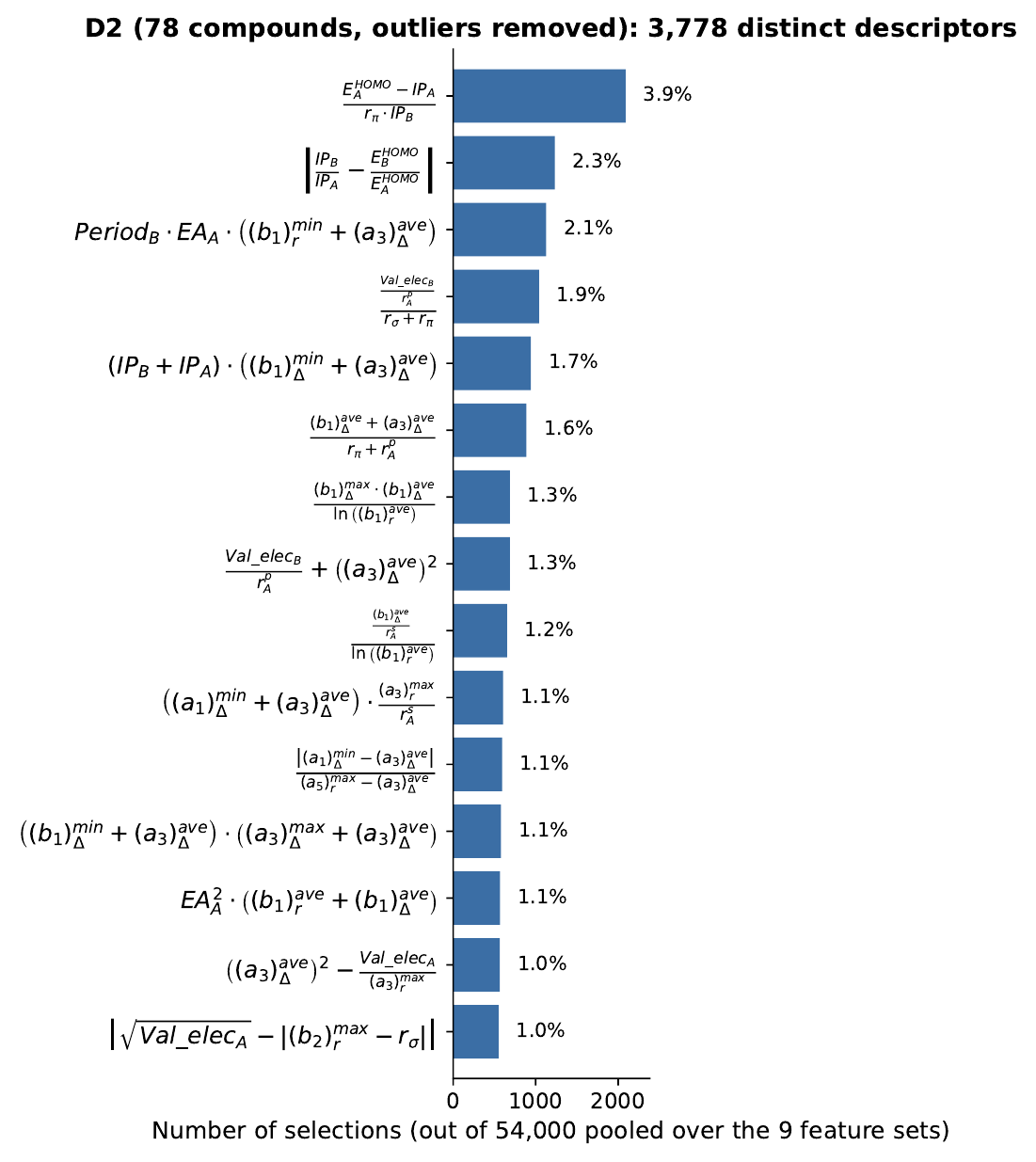}
    \caption{Globally most frequently selected 2D SISSO descriptor terms for dataset D2 (78 compounds, outliers removed); as in Figure~\ref{fig:sisso_desc_freq_D1} but for the refined dataset.}
    \label{fig:sisso_desc_freq_D2}
\end{figure}

\autoref{fig:sisso_desc_freq_D1} (D1) and \autoref{fig:sisso_desc_freq_D2} (D2) show the globally most frequently selected descriptor terms, pooled over all nine feature sets. The recursion features are written in the notation of the main text, $(a_n)^{\mathrm{agg}}_{r}$ and $(b_n)^{\mathrm{agg}}_{\Delta}$ for the ratio and the difference of the recursion coefficients between the RS and ZB structures, with the aggregation $\mathrm{agg}\in\{\mathrm{ave},\mathrm{min},\mathrm{max}\}$ over the two atoms.

\clearpage
\subsection{Feature importance of the SISSO ensemble}

\begin{figure}[htb]
    \centering
    \includegraphics[width=\textwidth]{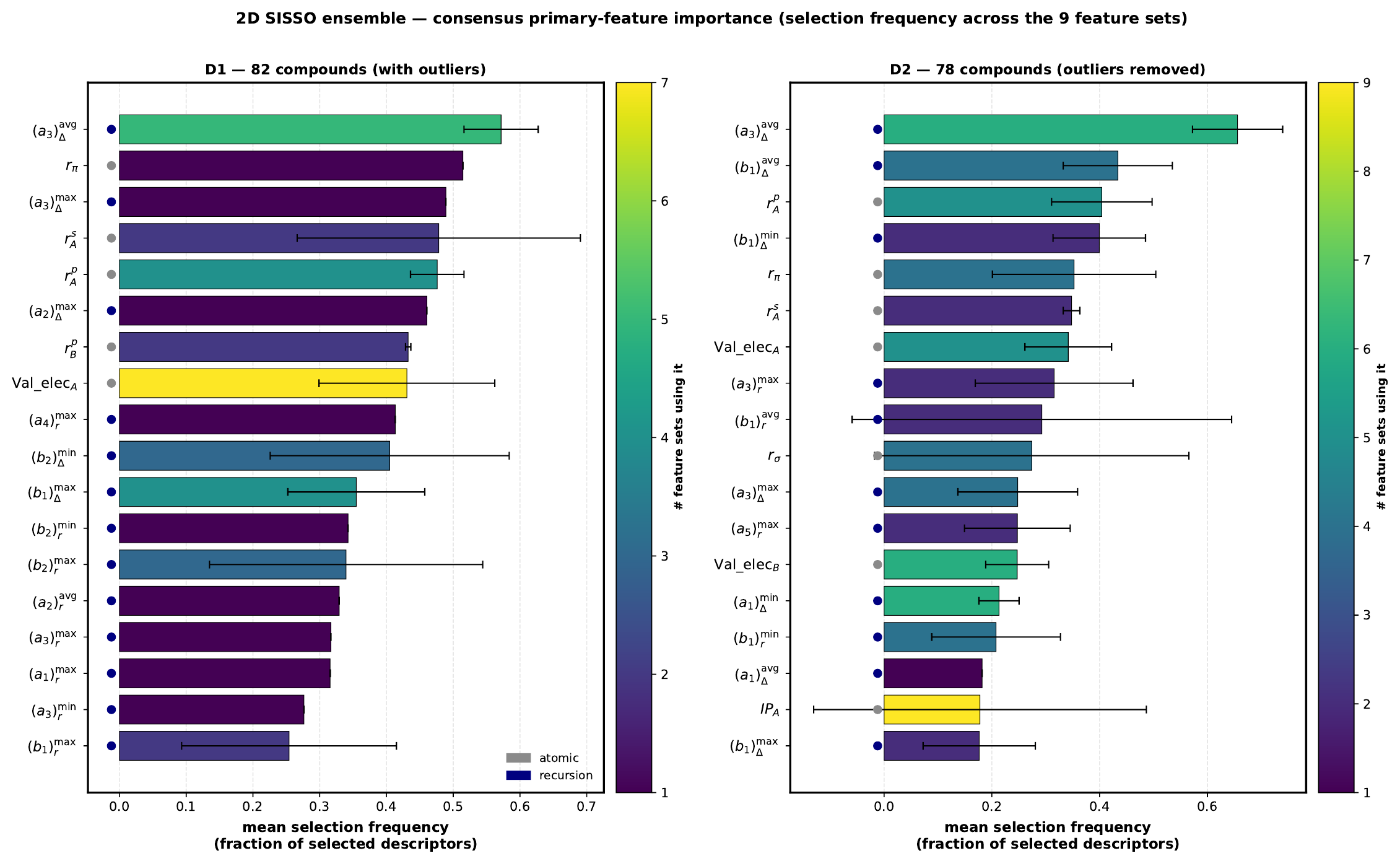}
    \caption{Selection frequency of the primary features in the 2D SISSO ensemble, averaged over the nine feature sets, for datasets D1 and D2. A feature counts as selected whenever it enters one of the two analytic terms of a rank-2 descriptor; the frequency is the share of the 6000 term selections per feature set in which it appears.}
    \label{fig:sisso_featimp}
\end{figure}

SISSO does not provide a permutation feature importance, because the descriptor is rebuilt for every member rather than evaluated on a fixed feature set. The natural importance measure is instead the \emph{selection frequency} of each primary feature, i.e.\ the fraction of the 6000 descriptor terms in which the feature appears. This is the same consensus quantity used for the KRR ensemble, so the two can be compared directly. \autoref{fig:sisso_featimp} shows the selection frequency of the primary features, averaged over the nine feature sets, for D1 and D2.

